\documentclass[preprint, amsmath,amssymb, aps,prd,showpacs]{revtex4-1}
\usepackage{epsfig}
\usepackage{dcolumn}
\usepackage{bm}
\usepackage{tikz}
\usepackage{graphicx, graphics, color}
\usepackage{dcolumn}
\usepackage{bm}
\usepackage{hyperref}
\usepackage[mathlines]{lineno}
\usepackage{float}
\usepackage{subfig}

\newcommand{\red}[1]{\textcolor[rgb]{0.00,0.00,0.00}{#1}}

\newcommand{\be}{\begin{equation}}
\newcommand{\ee}{\end{equation}}
\newcommand{\ben}{\begin{eqnarray}}
\newcommand{\een}{\end{eqnarray}}
\newcommand{\la}{{\lambda}}

\newcommand{\cO}{{\cal O}}

\newcommand{\na}{\nabla}

\newcommand{\ep}{\epsilon}

\newcommand{\ga}{\gamma}

\graphicspath{{figs/}}


\begin{document} 

\title{Static axionlike dark matter clouds around magnetized rotating wormholes - probe limit case}

\author{Bartlomiej Kiczek} 
\email{bkiczek@kft.umcs.lublin.pl}
\author{Marek Rogatko} 
\email{rogat@kft.umcs.lublin.pl}
\affiliation{Institute of Physics, 
Maria Curie-Sklodowska University, 
pl.~Marii Curie-Sklodowskiej 1,  20-031 Lublin,  Poland}

\date{\today}

\begin{abstract}
The problem of the distribution of axionlike particle, being the model of {\it dark matter}, in the nearby of rotating wormholes has been investigated numerically.
In the model in question the axion scalar is non-trivially coupled to the Maxwell gauge field. We consider two toy models of rotating wormholes embedded
in magnetic field, Kerr-like and Teo rotating wormholes. Moreover one assumes that the matter fields will not backreact on the wormhole spacetimes, i.e., we shall study the problem in the probe limit
case.
We point out the differences in the distribution of {\it dark matter} comparing to the location of it in the vicinity
of rotating magnetized black holes.
\end{abstract}

\maketitle
\flushbottom


\section{Introduction}
Recently there has been observed a big resurgence of interests in
a special class of Einstein field equation solutions representing tunnel-like structures connecting spatially separated regions or even more different Universes,
nowadays called wormholes.
\red{These fascinating objects are not only important for popular culture, but also gain a lot of scientific attention as their properties allow them to be black hole mimickers.}
From historical point of view,
the first description of such kind of objects begins with the issue of \cite{fla16}, devoted to spatial part of Schwarzschild solution studies.
The prototype of wormhole emerged from the studies devoted to particle model, where the mathematical construction which tried to eliminate 
coordinate or curvature singularities, dubbed as Einstein-Rosen bridge, was proposed in \cite{ein35}. 
Later on, the Kruskal-Szekeres coordinates were implemented for the description of Schwarzschild wormhole \cite{whe55}, 
while
the Euclidean form of wormhole solution was obtained in 
 \cite{haw88}. One should remark that all these concepts were postulated at quantum scale.

The current understanding of wormholes was revealed in \cite{mor88}, where the conditions for traversability for Lorentzian wormholes were defined by the survivability of human travellers.
\red{This redefinition was not only of great importance to physics, but also to futurology and is still seen as a main way to travel at large distances in space by humans.}
On the other hand, models of a wormhole, possessing no event horizon and physical singularities, were elaborated in 
 \cite{ell73}-\cite{ell79}. 
 In order to obtain such kind of wormhole solutions one should invoke phantom field (exotic matter), whose energy momentum tensor violates the null, weak and strong energy conditions, as well as, its kinetic energy term is of the reversed sign.
 
However, traversability requires also stability of the wormhole solution, except small acceleration and tidal forces. To achieve this goal we may consider 
a generalized
Einstein gravity theories,
like Gauss-Bonnet-dilaton theory. Moreover in this theory 
wormholes can be built with no use of such exotic kind of matter \cite{kan11}-\cite{har13}.  
On the other hand, the method of constructing traversable wormholes by applying duality rotation and complex transformations was proposed \cite{gib16,gib17}.
 By assuming that the dilaton field constitutes a phantom one, 
 an electrically charged traversable wormhole solution in Einstein-Maxwell-phantom dilaton gravity, has been revealed \cite{gou18}.

Soon after the rotating wormhole solutions were paid attention to \cite{teo98}-\cite{bro13}. 
There were also conceived perturbative and numerical attempts to construct spinning generalization of static wormhole solutions \cite{kas08}-\cite{che16}.
It was claimed that the rotating wormholes 
would be with a higher possibility stable \cite{mat06} and therefore traversable.

The other interesting problem in wormhole physics is their classification.
Having in mind classification delivered by the black hole uniqueness theorem,
the first work in this direction was provided in \cite{rub89}, delivering the uniqueness theorem for wormhole spaces with vanishing Ricci scalar. 
Further, the uniqueness of Ellis-Bronikov wormhole with phantom field was found in \cite{yaz17}, while the uniqueness
for four-dimensional case of the Einstein-Maxwell-dilaton wormholes with the dilaton coupling constant equal to one, was presented in \cite{laz17}.
The case of higher-dimensional generalization of wormhole solution, valid from the point of view of the unification theories like string/M-theory
attracts also attention.
The uniqueness theorem for higher-dimensional case of the static spherically symmetric phantom wormholes was treated in \cite{rog18}, while
the case of of static spherically symmetric traversable wormholes with two asymptotically flat ends, subject to the higher-dimensional solutions
of Einstein-Maxwell-phantom dilaton field equations with an arbitrary dilaton coupling constant, was elaborated in \cite{rog18a}.

Various other aspects of physics of these objects were under intensive studies (for a detailed review of the blossoming subject the reader may consult \cite{worm}).

Wormholes being a fascinating subject of their possible impact on space and time travels, may also be regarded as potential
astrophysical objects, that can be observationally search for.
From the astrophysical point of view, it is persuasive to consider rotating wormholes. The problem that arises is how to observationally distinguish
rotating wormholes from stationary axisymmetric black holes of Kerr-type. Remarkable attention to the aforementioned problem was paid to after
the Even Horizon Telescope observed the black hole shadow in the center of the galaxy M87.\\
The first studies to what extent wormholes can imitate the observational characteristics of black holes were conducted in \cite{dam07},
where the simple generalization of Schwarzschild-like line element was revealed.
The considered metric differs from the static general relativity one by introducing the dimensionless parameter $\la$. The value of the parameter equal to zero is 
responsible for the ordinary Schwarzschild black hole solution.

Of course one should be aware that for non-zero values of the parameter the presented line element is no longer the static solution of Einstein equations
and changes the structure of the manifold. Therefore the matter with almost vanishing energy density ought to be required to maintain the aforementioned
gravitational configuration (for the discussion of the influence of the parameter $\la$ on the static manifold structure see, e.g., \cite{bue18}).

Further generalization of the idea given in \cite{dam07} to describe Kerr-like wormhole spacetime as a toy model, was achieved by applying
a modification on the Kerr metric similar to the procedure performed in \cite{dam07}.  
The embedding diagrams, geodesic structure, as well as, shadow characteristics of the obtained Kerr-like wormhole were given in \cite{ami19}.
On the other hand, the throat-like effects on the shadow of Kerr-like wormholes were elaborated in \cite{kas21}.

However,
the problem of the structure at the horizon scale of black hole which gives rise to echoes of the gravitational wave signal bounded with the postmeger 
ring-down phase in binary coalescences, in the case of static and rotating toy models of traversable wormholes, has been
elucidated in \cite{bue18}.

The other subject acquiring much attention in contemporary astrophysics and physics is the unrelenting search for finding {\it dark matter}
sector particles. The nature of this elusive ingredient of our Universe is a mystery and several models try to explain it and constitute the possible guidance for the future experiments. 
The main aim of our work will be to investigate the behavior of axion-like particle {\it dark matter} model clouds, around the mimickers of rotating 
black holes, stationary axially symmetric wormholes. The work will provide some continuity with our previous studies \cite{kic21}, where we have paid
attention to the main features of axionic clouds {\it dark matter} in the vicinity of magnetized rotating black holes.

The principal goal of the investigations will be to find the possible differences in characteristic features of the axion-like condensate, between those two classes of compact objects,
i.e., rotating black holes and black hole mimickers. Our studies will constitute the first glimpse at the problem in question. Namely, we restrict our consideration to the probe limit case,
when one has the complete separation of the degrees of freedom, i.e., matter fields do not backreact on wormhole spacetime.

The organization of the paper is as follows. In Sec. II we deliver the basic facts about the axion-like {\it dark matter} model. Sec. III 
will be devoted to the description of the rotating wormholes models surrounded by {\it dark matter} clouds, in the considered model
of axion-like {\it dark matter}. In Sec. IV we describe the numerical results of the studies, while in Sec. V we conclude our investigations and aim the possible
problems for the future investigations.

\section{Model of axion-like {\it dark matter} sector}

The explanation of astronomical and cosmological observations require {\it dark matter} existence, whose nature is one of the most tantalizing questions confronting 
contemporary physics and cosmology. A large number of ongoing or planned experimental searches for its detection and understanding of the {\it dark sector} role
in a fundamental description of the Universe.
Axions are among the strongest candidates for the possible explanation of the existence of {\it hidden sector} \cite{pre83}-\cite{din83}.
Their existence has been postulated to explain the apparent lack of violation of charge conjugate parity \cite{pec77}-\cite{wil78} and in the strong interaction
motivated the absence of observable electric dipole moment of the neutron \cite{pen15}. Axionlike particles are also widely spotted in the realm of string theories \cite{svr06}.

In what follows, we shall study axionlike scalar particles coupled to the 
Maxwell $U(1)$-gauge field. The non-trivial coupling of axion field to the Maxwell field strength invariant plays the crucial role in the model in question.
The field equations of motion are provided by the variation procedure with respect to the action given by
\begin{equation}
\mathcal{S} = \int d^4 x \sqrt{-g} \left[R - \frac{1}{4} F_{\mu \nu} F^{\mu \nu}
 - \frac{1}{2} \nabla_\mu \Psi \nabla^\mu \Psi - \frac{\mu^2}{2} \Psi^2
 - \frac{k}{2} \Psi \ast F^{\mu \nu} F_{\mu \nu} \right],
\end{equation}
where we set $R$ for the Ricci scalar, $F_{\mu \nu} = 2 \nabla_{[\mu} A_{\nu]}$, while $\Psi$ stands for the scalar field  (axion) with mass $\mu$.
$\ast F^{\mu \nu} = 1/2 \ep_{\mu \nu \alpha \beta} F^{\alpha \beta}$ is the dual to Maxwell field strength.

The equation of motion for the scalar field $\Psi$, which constitutes a covariant Klein-Gordon equation with a source term of the dual Maxwell field invariant, implies
\begin{equation}
\nabla_\mu \nabla^\mu \Psi - \mu^2 \Psi 
- \frac{k}{2} ~\ast F^{\mu \nu} F_{\mu \nu}  = 0,
\label{eq:field_eqn}
\end{equation}
while the $U(1)$-gauge field is subject to the relation as follows:
\be
\na_\mu F^{\nu \mu} + 2 k~\ast F^{\nu \mu} \na_{\mu }\Psi = 0.
\ee
We refer to the $\Psi$ field as axionlike, because the axions (originating from QCD) have adequate constrains on both mass and coupling parameter. Here however we consider particles with physics given by an analogical Lagrangian yet with arbitrary values of physical parameters. However for simplicity we might refer to the studied axionlike particles as simply axions.

The {\it dark matter} model in question was widely elaborated in studies of black hole superradiance
and light polarization effects, possible experimental signals of {\it dark sector} around these objects \cite{pla18}-\cite{car18}, \cite{kic21}, 
and neutron stars \cite{gar18}-\cite{gra15}, as well as,
the influence of axionic {\it dark matter} on the physics on early Universe and primordial black holes \cite{fed19}-\cite{ros18}.

The form of the relation (\ref{eq:field_eqn}) envisages the fact that the presence of the non-zero source term, containing the dual invariant, given by
\begin{equation}
\mathcal{I} = ~\ast F^{\mu \nu} F_{\mu \nu} \neq 0,
\end{equation}
is crucial.
In the opposite case, when the invariant is equal to zero, the axion-like scalar field equation of motion reduces to the simple massive Klein-Gordon case,
without any self-interaction potential. It means that  no scalar hair configuration on the studied line element can emerge.
Although it has been shown that in Kerr spacetime scalar hair may emerge in certain situations \cite{herd14}, here we pick a different ansatz (see below) as we focus on stationary configurations, which appear to be magnetically induced in this approach.
On the other hand, it can be noticed that the discussed invariant, $\ast F_{\mu \nu} F^{\mu \nu}$,
is equal to zero in the case when $F_{\mu \nu} =0$, or for spherically symmetric spacetime. However, it has a non-trivial form, $\ast F_{\mu \nu} F^{\mu \nu} \neq 0$, 
 when both rotation and magnetic $U(1)$-gauge field components are present in the spacetime under consideration.

To introduce the magnetic field we use the method proposed by Wald \cite{wal74}, where the vector potential is sourced by Killing vectors of the rotating spacetime.
In general it has a form
\begin{equation}
A_\mu = \frac{1}{2}B (m_\mu + 2 a k_\mu),
\end{equation}
where $k_\mu$ and $m_\mu$ are the Killing vectors connected with temporal invariance and  $\phi$ rotation respectively.

As in \cite{kic21}, where we have studied rotating magnetized black holes submerged into axionic {\it dark matter} cloud, one can introduce a static magnetic
field to the system, which will be oriented along the rotation axis. It seems to be plausible from the point of astrophysical perspective
and can be regarded as a starting point for studies of the magnetic field influence of the system in question. Because of the fact that our investigations
focus on static magnetic field, parallel to the wormhole rotation axis, the gauge potential may be rewritten in the form as $
A_\mu dx^\mu = B/2~ g_{\mu \nu} m^\nu dx^\mu.$

For our considerations we choose a static, time independent ansatz. The symmetry of the problem enables us to elaborate
the axion field in the form provided by
\begin{equation}
\Psi = \psi(r, \theta),
\label{eq:ansatz}
\end{equation}
which will be plugged into the equation \eqref{eq:field_eqn}, for the considered line element.

\section{Rotating wormhole metrics}
The simplicity of the static line element describing a wormhole may suggest that the spinning generalization can be achieved analytically 
and ought to be globally regular. But in vain, it happens that finding the stationary solution with an extended source is far more complicated (see
for the recent aspects of this problem \cite{vol21}). However, the rotating wormhole solutions are widely discussed in literature \cite{teo98}-\cite{che16},
but one should be aware that they do not constitute the exact solutions of the equations of motion but rather comprise some model of geometries.

In this section, we shall study two kinds of rotating wormhole model metrics. First one accounts for the extension of the regular black hole Kerr metric \cite{bue18,ami19}. The other is the Teo class wormhole \cite{teo98},  a rotating generalization of Morris-Thorne wormhole, which serves us as comparison to a bit more realistic Kerr-like wormhole.

\subsection{Kerr-like wormhole}
To begin with, we consider the metric of Kerr-like rotating wormhole.
It is constructed by a slight modification of stationary axisymmetric line element with a parameter $\la$.
For the first time, such construction was proposed in  \cite{dam07}, where the static Schwarzschild black hole was considered.
Then, it was generalized to the case of stationary axisymmetric line element \cite{bue18,ami19}.
The Kerr-like wormhole line element yields
\ben
ds^2 &=& - \left( 1 - \frac{2 M r}{\Sigma} \right)dt^2 - \frac{4 M ar \sin^2 \theta}{\Sigma} dt d\phi + \frac{\Sigma}{\tilde{\Delta}} dr^2 + \Sigma d\theta^2\\ \nonumber
&+& \Big(r^2 + a^2 + \frac{2 M a^2 r \sin^2 \theta}{\Sigma} \Big) \sin^2 \theta d\phi^2,
\een
where we set 
\begin{align}
\Sigma(r, \theta) = r^2 + a^2 cos^2 \theta, \\
\tilde{\Delta}(r) = r^2 + a^2 - 2M(1 + \lambda^2)r.
\end{align}
The parameters $M$ and $a M$ correspond to mass and angular momentum of a wormhole. 
For a small deviation parameter $\la$, one achieves
almost indistinguishable from of Kerr black hole line element. These three parameters describe the system as seen from the outside.
Moreover its Arnowitt-Deser-Misner (ADM) mass, as seen by the observer at asymptotic spatial infinity, is given by $M_{ADM} = M (1 + \la^2)$.

The largest root of $\tilde{\Delta}(r) = 0$, establishes the surface provided by 
\begin{equation}
r_+ = M ( 1 + \lambda^2 ) + \sqrt{M^2 ( 1 + \lambda^2)^2 - a^2}.
\end{equation}
For the model in question it does not constitute a radius of the event horizon, but describes the radius of the throat of the rotating wormhole, which 
connects two asymptotically flat regions of the spacetime. It can be explicitly seen by the adequate changes of variables \cite{bue18,ami19}. The points
with the condition $r<r_+$ do not exist. 

Consequently the axion field equation written in the Kerr-like wormhole spacetime implies the following:
\begin{align}
\tilde{\Delta} \partial_r^2 \psi + 
\frac{2(r - M)\tilde{\Delta} - M \lambda^2 (r^2 + a^2)}{\Delta} \partial_r \psi
+ \partial_{\theta}^2 \psi
+  \cot \theta \partial_{\theta} \psi
- \mu^2 \Sigma \psi = \frac{k \Sigma}{2} \mathcal{I}_{KWH},
\label{eqn:kwh_axion}
\end{align}
where 
the electromagnetic field invariant is provided by
\begin{align}
\mathcal{I}_{KWH} = - \frac{a B^2 M \tilde{\Delta} \sin^2 \theta \cos \theta}{2 \Delta \Sigma^4} \big[ 3 a^6 + 2 a^4 M r - 5 a^4 r^2 - 8 a^2 M r^3 - 32 a^2 r^4 - 24 r^6 \nonumber \\ 
+ 4 a^2 (a^4 - a^2 r^2 + 2(M - r)r^3 ) \cos 2\theta + a^4 (a^2 - 2 M r + r^2) \cos 4\theta \big]. 
\end{align}
The equation \eqref{eqn:kwh_axion} undergoes a following scaling transformation
\begin{equation}
r \rightarrow \eta r, \quad a \rightarrow \eta a, \quad M \rightarrow \eta M, \quad
B \rightarrow B/\eta, \quad \mu^2 \rightarrow \mu^2 / \eta^2,  \quad r_+ \rightarrow \eta r_+,
\end{equation}
\red{which allows us to fix one of model parameters to unity. For this we pick $M = 1$.}

\subsection{Teo rotating wormhole}
The well-known Morris-Thorne metric, introduced in Ref. \cite{mor88}, 
describes a traversable wormhole spacetime, which is stabilised by exotic matter in the area of its throat.
That solution was achieved by using reverse engineering of general relativity, namely the metric was postulated first and with a help of Einstein equations 
the suitable matter components were found.
Generalization of the aforementioned solution, by including the rotation into the consideration, was performed in \cite{teo98}.
The resulting metric of the rotating wormhole has a following form:
\begin{equation}
ds^2 =  -N^2 dt^2 + \frac{dr^2}{1 - \frac{b}{r}} + K^2 r^2 \left[ d \theta^2 + \sin^2 \theta (d \phi - \omega dt)^2 \right],
\end{equation}
where, as in the Morris-Thorne case,  one has a lot of freedom in choosing the shape of $N$, $b$, $K$ and $\omega$ functions, as long as they meet specific requirements.
Firstly, all the functions can be functions of $r$ and $\theta$ and should be regular on the symmetry axis $\theta =0, \pi$.
Secondly, $N$, the gravitational redshift function, ought to be finite and nonzero, $b$ as the shape function determining the shape of the wormhole throat,
should satisfy $b \leqslant r$.
$K$ accounts for the radial distance with respect to the coordinate origin and $\omega$ stands for the angular velocity of the wormhole.

The embedding of constant $t$ and $\theta$-cross sections in the three-dimensional Euclidean space reveals the well-recognizable form of the wormhole
spacetime. The constructed geometry describes two regions, where the radial coordinates are given by $r \in [r_+,~\infty)$, which are joined together
 at the wormhole throat $r=r_+$. At spatial infinity, the requirement of asymptotic flatness regions provides that the metric coefficients ought to satisfy
 the following expansions:
 \be
 N = 1 - \frac{M}{r} + \cO \Big(\frac{1}{r^2}\Big), \qquad K = 1 + \cO\Big(\frac{1}{r}\Big),
 \qquad \frac{b}{r} = \cO\Big(\frac{1}{r}\Big), \qquad \omega = \frac{2 J}{r^3} + \cO\Big(\frac{1}{r^4}\Big),
 \label{eq:twh_asympt}
 \ee
 where we have denoted by $M$ the mass of the wormhole and by $J$ its angular momentum.
 In general, one encounters the whole range of functions, which fulfil the aforementioned conditions and constitute a regular rotating wormhole solution.

For the numerical calculations, we pick a set of functions which appear to be quite popular in the literature of the subject,
and were previously used by different authors \cite{shaikh18, nedkova13, abdujabbarov16, harko09, bambi13}
\be
N = \exp\left[- \frac{r_+}{r} \right], \qquad b(r) = r_+ \left( \frac{r_+}{r} \right)^\gamma, \qquad
\omega = \frac{2 a r_+}{r^3}, \qquad K=1,
\label{eq:twh_metric_fun}
\ee
where we use the $r_+$ symbol, for denoting the wormhole throat radius. 
\red{The angular momentum parameter is defined in the standard way $a = J/M$. 
Using the asymptotic relations \eqref{eq:twh_asympt} we find that for the picked set of functions \eqref{eq:twh_metric_fun} $M = r_+$.}
Thus, the family of the above
solutions is described by three parameters, i.e., the throat radius $r_+$, angular momentum parameter $a$ and the shape parameter $\gamma$.

After putting the ansatz \eqref{eq:ansatz} and the metric into the field equation \eqref{eq:axion_only_action} we arrive at the equation of motion
\begin{align}
\left[ r^2 - r_+ r \left( \frac{r_+}{r} \right)^\gamma \right] \partial_r^2 \psi + \left[ 2r + r_+ + \left(\frac{r_+}{r} \right)^\gamma \left(\frac{1}{2}r_+ \gamma - \frac{r_+^2}{r} -\frac{3}{2} r_+ \right) \right] \partial_r \psi \nonumber \\ 
+ \partial_{\theta}^2 \psi + \cot \theta \partial_{\theta} \psi - \mu^2 r^2 \psi = \frac{1}{2} k r^2 \mathcal{I}_{TWH},
\label{eqn:twh_axion}
\end{align}
which radial part is strongly dependent on $\gamma$.
The Maxwell field invariant related to uniform magnetic field in this spacetime implies
\begin{equation}
\mathcal{I}_{TWH} = \frac{12 a B^2 r_+ \cos \theta \sin^2 \theta}{r^{5/2}} \sqrt{\frac{r - r_+ \left(\frac{r_+}{r} \right)^\gamma}{\exp \left[ -\frac{2 r_+}{r} \right]}}.
\end{equation}
The equation \eqref{eqn:twh_axion} follows a scaling transformation
\begin{equation}
r \rightarrow \eta r, \quad r_+ \rightarrow \eta r_+, \quad a \rightarrow \eta a, \quad B \rightarrow B/ \eta, \quad \mu^2 \rightarrow \mu^2 / \eta^2.
\end{equation}
\red{Using this transformation we fix $r_+ = 1$.}

\subsection{Free energy}
As a benchmark for the thermodynamical preference of the obtained states we use free energy by evaluating the on-shell action of the axion dependent part of the theory
\begin{equation}
\mathcal{S}_{axion} = \int d^4 x \sqrt{-g} \left[- \frac{1}{2} \nabla_\mu \Psi \nabla^\mu \Psi - \frac{\mu^2}{2} \Psi^2
 - \frac{k}{2} \Psi  \ast F^{\mu \nu} F_{\mu \nu} \right].
\label{eq:axion_only_action}
\end{equation}
By substituting the equations of motion into the action and imposing the ansatz of the field we arrive to the formula for the free energy
\begin{equation}
F = - 2 \pi \int_\mathcal{M} dr d\theta ~\sqrt{-g} \bigg[ (\partial_r \psi)^2 g^{rr} + (\partial_\theta \psi)^2 g^{\theta \theta} + \mu^2 \psi^2 \bigg].
\label{eq_freeenergy}
\end{equation}
The straightforward integration of the equation \eqref{eq_freeenergy} appears to be problematic.  It is because both considered backgrounds have singular metric determinant at the throat, which makes simple integration from throat to infinity impossible in these coordinates.
It should be noted that this singularity is merely a coordinate singularity, as the curvature of both wormholes is regular and finite at the throat.

In 
the case of Kerr-like wormhole metric,
 we have
\be
\sqrt{-g} = \sqrt{\frac{\Delta}{\tilde{\Delta}}} \Sigma \sin^2 \theta,
\ee
where for the case of $\la$ equal to zero we obtain that $\Delta = {\tilde{\Delta}}$.
This fact
naturally eradicates the singularity problem in the black hole scenario. Here, however, as we radially fall toward the wormhole, the root of ${\tilde{\Delta}}$ comes first and creates the singularity.
On the other hand, for the Teo rotating line element we get
\begin{equation}
\sqrt{-g} = \frac{\exp \left( - \frac{r_+}{r} \right) r^2 \sin \theta}{\sqrt{1 - \left( \frac{r_+}{r} \right)^{\gamma + 1}}},
\end{equation}
with the denominator naturally generating the infinity.

To deal with the integration in such spacetimes we use energy differences instead.
Also we introduce a cutoff to the lower integration bound, so we start from $r_+ + \epsilon$ rather than simply $r_+$.
In this way we ensure the finiteness of energy differences and give them straightforward physical interpretation.
With the change of the background parameter the solution becomes more or less thermodynamically stable with respect to some \textit{ground} solution.

\section{Results}
In this section we pay attention to the solutions of the equations of motion for the previously described two toy models of rotating wormholes.
Due to the complications of the relations \eqref{eqn:kwh_axion} and \eqref{eqn:twh_axion},
we solve them numerically by virtue of spectral methods. Firstly the adequate equation is discretized on Gauss-Lobato grid \cite{matlabnum} and 
next translated into a system of algebraic equations with spectral differentiation matrices.
The method in question 
has already been implemented in Python and tested for the numerical stability.
The technical details, especially convergence tests of the numerical method are described in the Appendix of \cite{kic21}, where we studied
the problem of axionlike particle clouds in the spacetimes of rotating magnetized black holes.

The spectral nature of the numerical scheme requires remapping the coordinates onto the $[-1, 1]$ intervals.
It can be achieved by the coordinate transformation provided by
\begin{align}
z = 1 - \frac{2 r_+}{r}, \\ 
u = \frac{4 \theta}{\pi} - 1,
\end{align}
where $r_+$ is the wormhole throat radius.
After such operation, our numerical domain may be written in the form $[-1, 1]\times[-1, 1]$.
For $z$-coordinate, the boundaries are the wormhole throat ($z = -1$) and spatial infinity ($z=1$), while for $u = -1$, one talks about \textit{north pole} of a wormhole and the \textit{equator} with $u = 1$.

Consequently after the coordinate transformation in the underlying equations, one shall impose the adequate boundary conditions.
Namely, on the throat surface we demand that the axion field should be regular, therefore $\partial_r \psi = 0$ provides a desirable conduct of the field.
\red{Alternatively, setting the field to a constant value, such as zero in a wormhole scenario, is also a possible choice. However we wish to explore the Kerr-like solution for different values of $\lambda$ parameter, including its zeroing when it simplifies to the Kerr black hole.
Given that for the consistency between these two kinds of solutions we use the Neumann boundary condition.}
At the spatial infinity,  we take a look on the asymptotic behaviour of the equation itself and the source term $\mathcal{I}$.
It appears that the Maxwell field invariants in both backgrounds are vanishing functions. As $r \rightarrow \infty$, we have
\begin{equation}
I_{KWH} = \mathcal{O}\left(\frac{1}{r^4}\right),
\end{equation}
\begin{equation}
I_{TWH} = \mathcal{O}\left( \frac{1}{r^2} \right).
\end{equation}
Which means that both equations \eqref{eqn:kwh_axion} and \eqref{eqn:twh_axion} reach a simple, asymptotic form, to the leading order
\begin{equation}
\partial^2_r \psi + \frac{2}{r} \partial_r \psi - \mu^2 \psi = 0.
\end{equation}
This simple equation has a solution
\begin{equation}
\psi = A \frac{\exp(\mu r)}{r} + B \frac{\exp(-\mu r)}{r},
\label{eq:psi_asympt}
\end{equation}
where $A$ and $B$ are constants.
Naturally the field ought to decay for the sake of asymptotic flatness of the spacetime.
Given that we are allowed to choose $A = 0$, with arbitrary $B$.
This means that a boundary condition $\psi(r \rightarrow \infty) = 0$ is an adequate and mathematically motivated choice.

On the other hand, the boundary conditions for the angular dependency are built on the basis of the spacetime symmetry. Both considered spacetimes are rotating, therefore we demand $\partial_{\theta} \psi = 0$ on the \textit{north pole}.
On the \textit{equator}, the presence of magnetic field combined with the spacetime symmetry implies that $\psi = 0$.

\subsection{Kerr-like wormhole}
\begin{figure}[h]
\centering
\subfloat[$\mu^2 \rightarrow 0^+, \quad \lambda = 0.5$]{
\includegraphics[width=0.45 \textwidth]{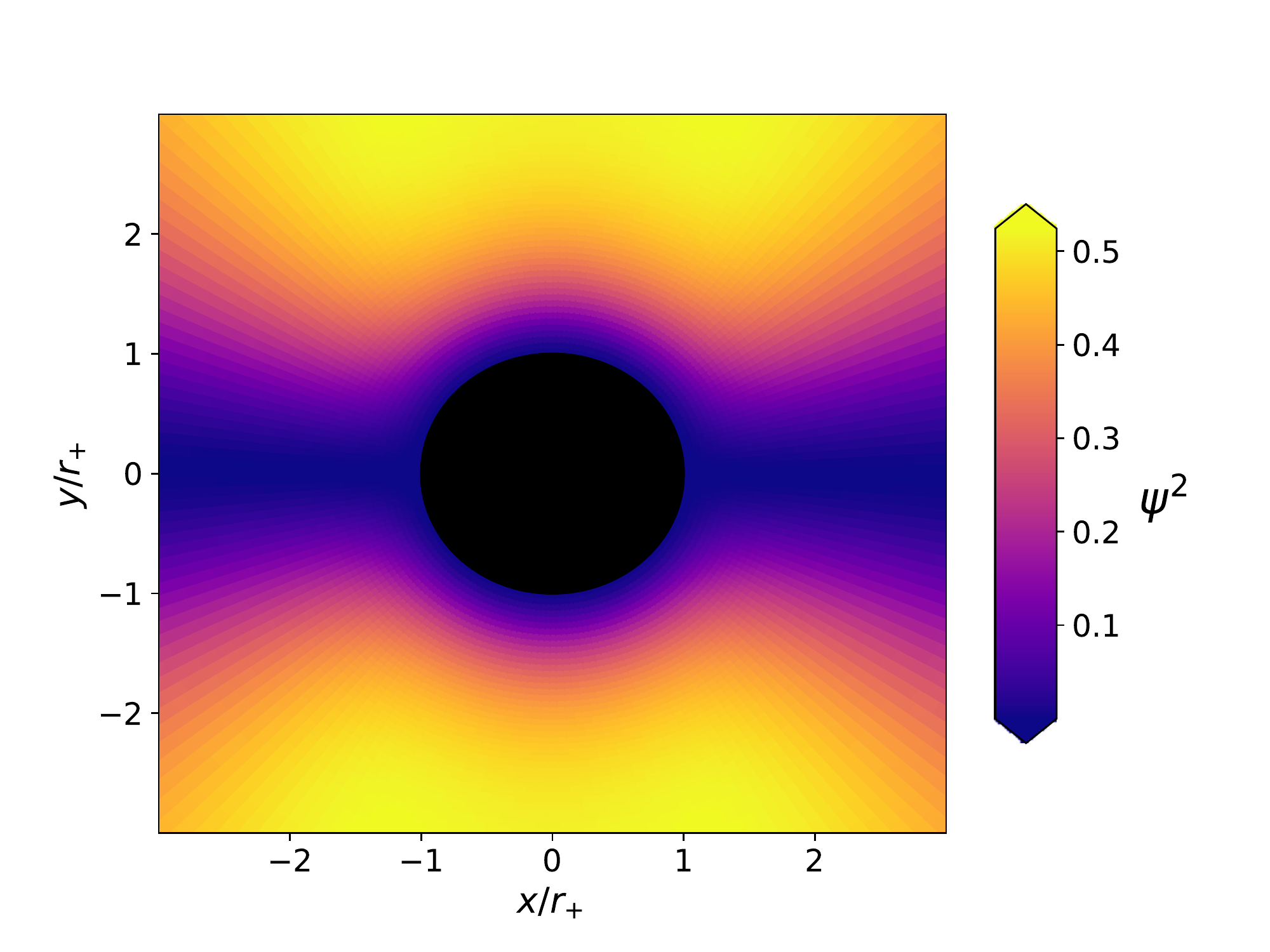}
}
\qquad
\subfloat[$\mu^2 = 0.01, \quad \lambda = 0.5$]{
\includegraphics[width=0.45 \textwidth]{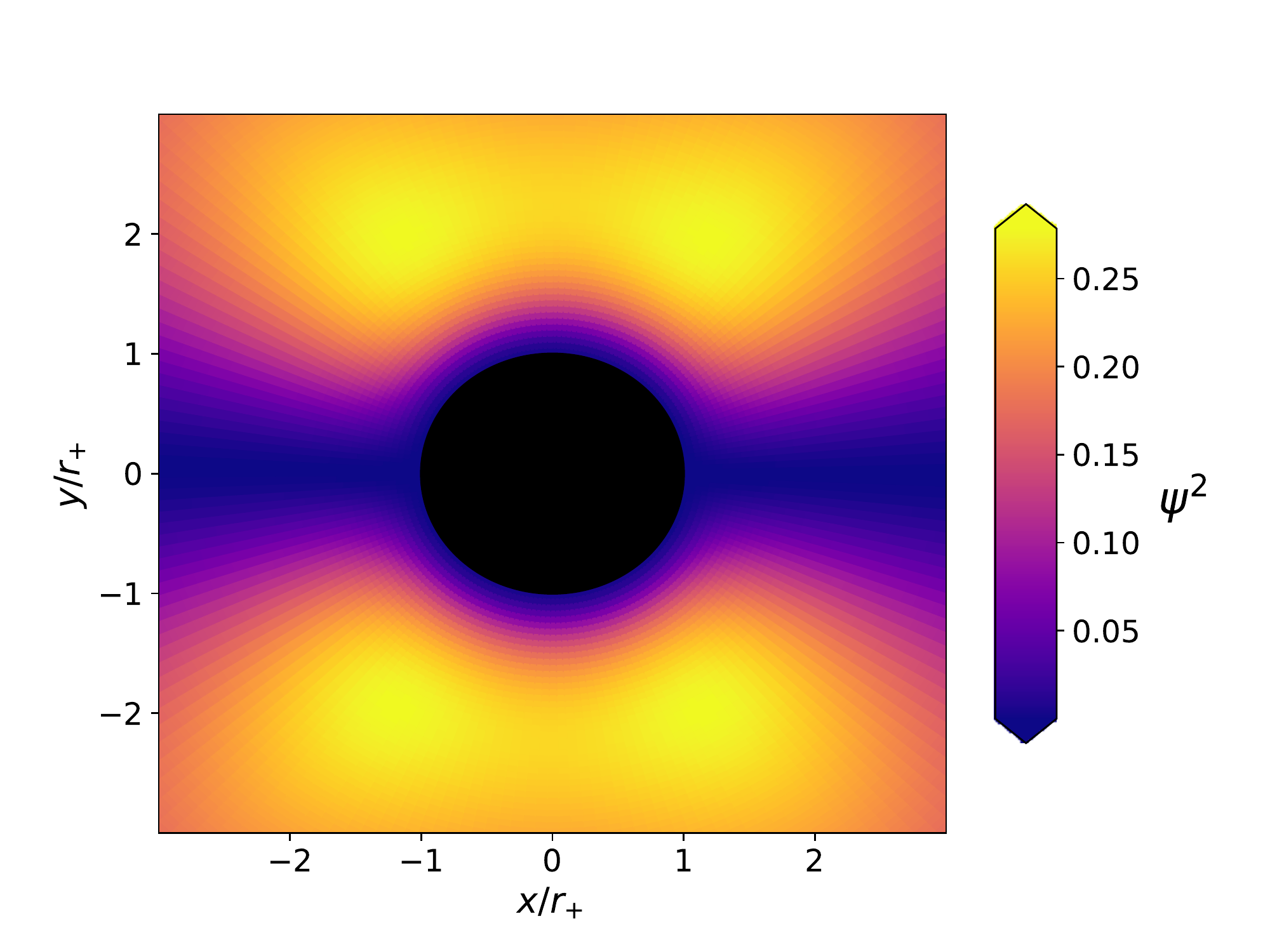}
}

\vspace{0.5cm}
\subfloat[$\mu^2 = 0.1, \quad  \lambda = 0.5$]{
\includegraphics[width=0.45 \textwidth]{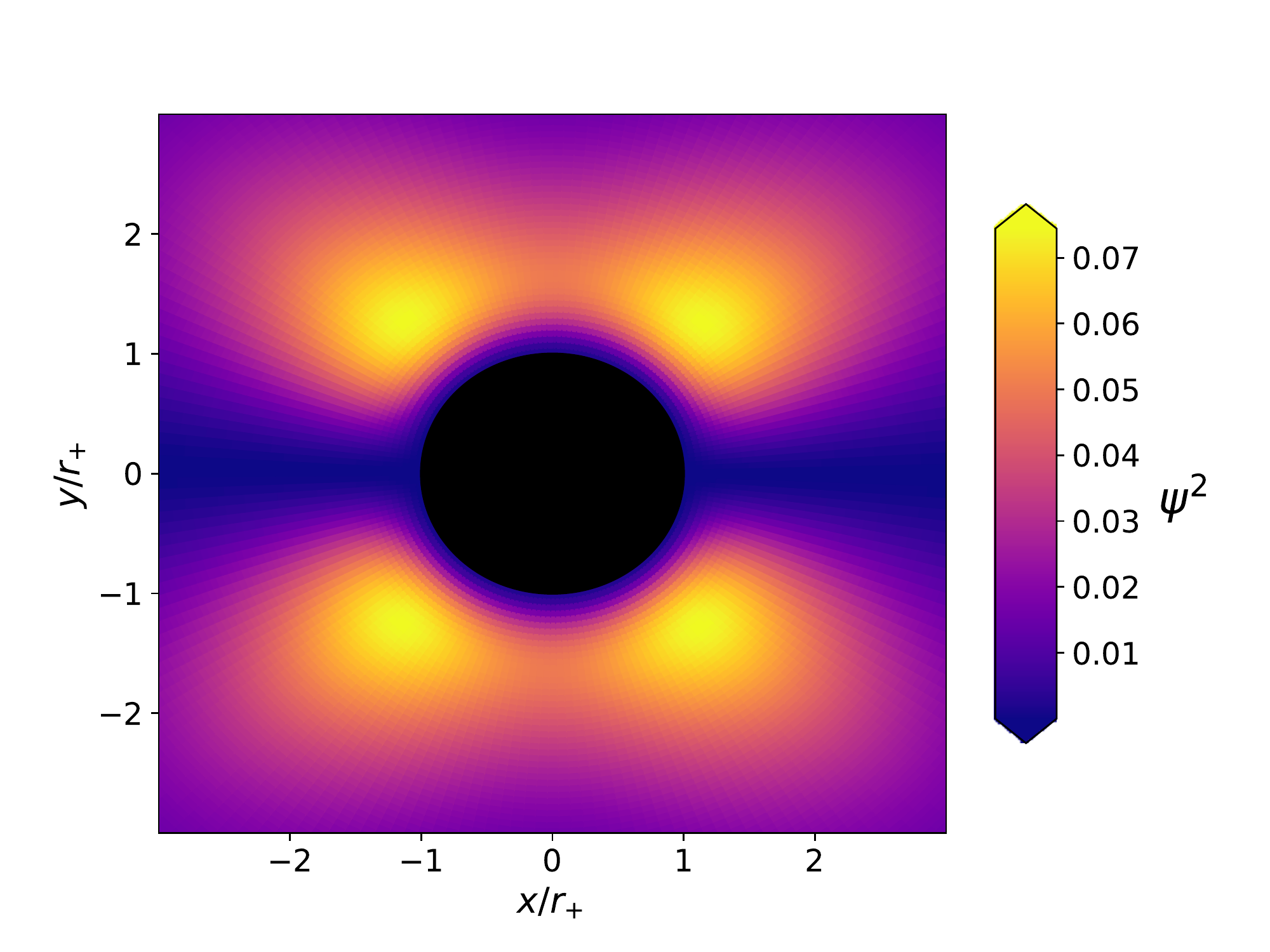}
}
\qquad
\subfloat[$\mu^2 = 1, \quad \lambda = 0.5$]{
\includegraphics[width=0.45 \textwidth]{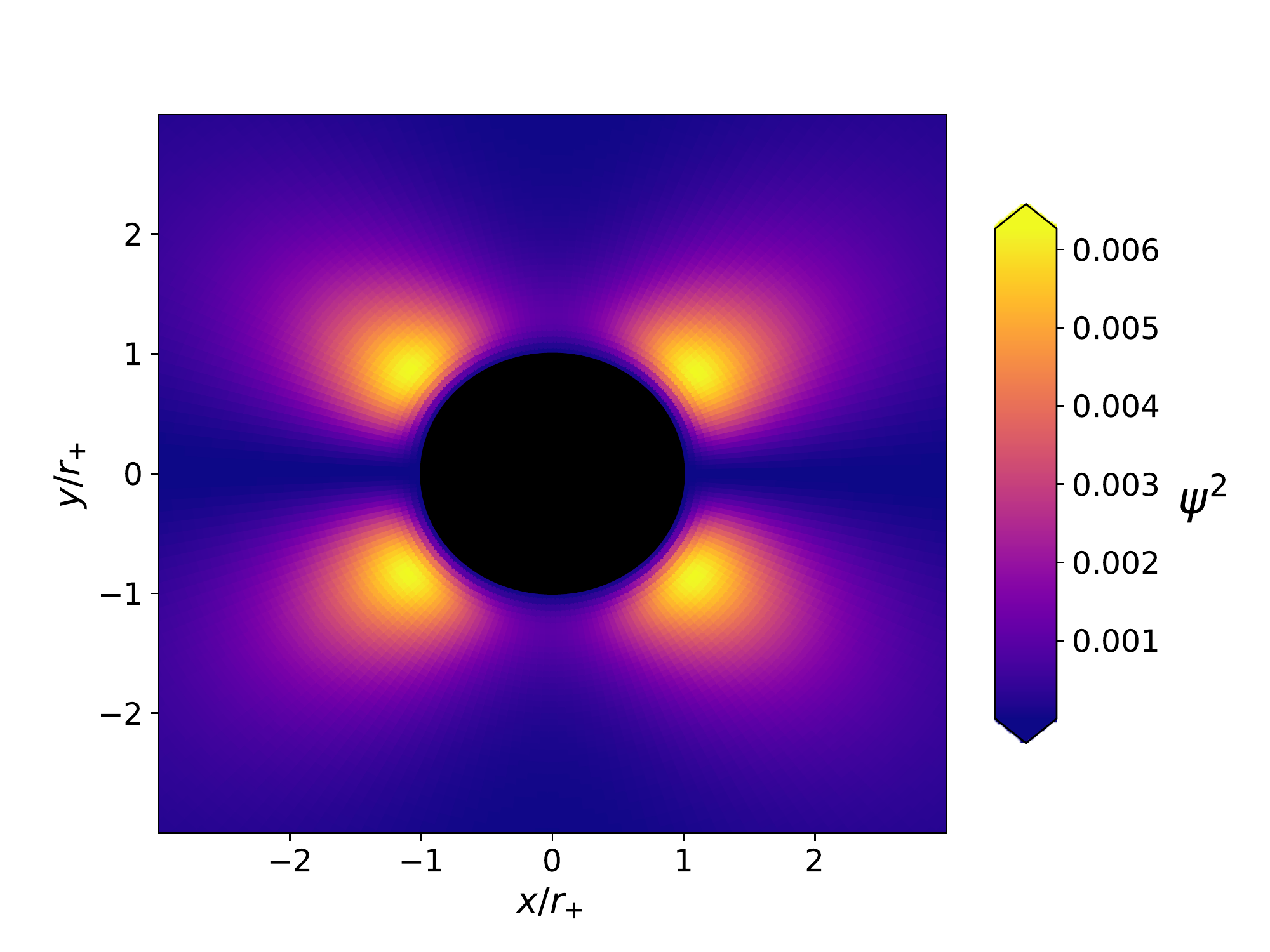}
}
\caption{Axion field distribution around Kerr-like wormholes for given sets of parameters.
The blank space in the vicinity of wormhole throat distinguishes the solution from the black hole counterpart, where the field appears to be non-zero on the event horizon.
Subsequent panels for each mass parameter show how the angular distribution of the field is affected.}
\label{fig_kw_maps}
\end{figure}

To commence with, we solve the equation \eqref{eqn:kwh_axion} for the Kerr-like background metric.
A portion of obtained distributions is depicted in Fig. \ref{fig_kw_maps}.
In the following panels we see the increasing mass of the axionic field.
In the panel (a) the field is ultralight, subsequently in (b) $\mu^2 = 0.01$,  (c) $\mu^2 = 0.1$ and finally in (d) $\mu^2 = 1$.
In every panel we have $a = 0.99$ and $\lambda = 0.5$. We can clearly see how the mass of the axionlike field changes the angular distribution of it around the wormhole.
For little masses the clouds are concentrated around the poles of the wormhole and spread in the space for several throat radii.
As we increase the axion mass we see that the polar regions of the wormhole become depleted and the field drifts towards the equator. The largest concentration is visible on the latitude $\theta \simeq \pi/4$.

Second important effect is the influence of the field mass on the magnitude of the field.
Inspection of
the colorbars reveals that the larger the mass the smaller the field.
The spatial tail of the field is also much shorter, when the mass of the field is larger.
Intuitively, in the asymptotic solution \eqref{eq:psi_asympt} $\mu$ enters the suppressing exponential term.  The field decays faster for larger masses, which means the massive fields are localized in the vicinity of the throat surface.

Another important thing that stands out in relation to the black hole solutions is the repulsion of the axion cloud from the wormhole throat surface.
While in the case of the black hole, the field had non-zero values on the surface of the event horizon,  and its radial character was monotonically decreasing, here we have a completely different situation.
For the wormhole, the field vanishes or at least has a significantly smaller value on the throat.
Then it grows with the radius as it reaches the maximum and finally decreases.
This effect is particularly visible for the high values of the angular momentum.

\begin{figure}
\centering
\includegraphics[width=0.8\textwidth]{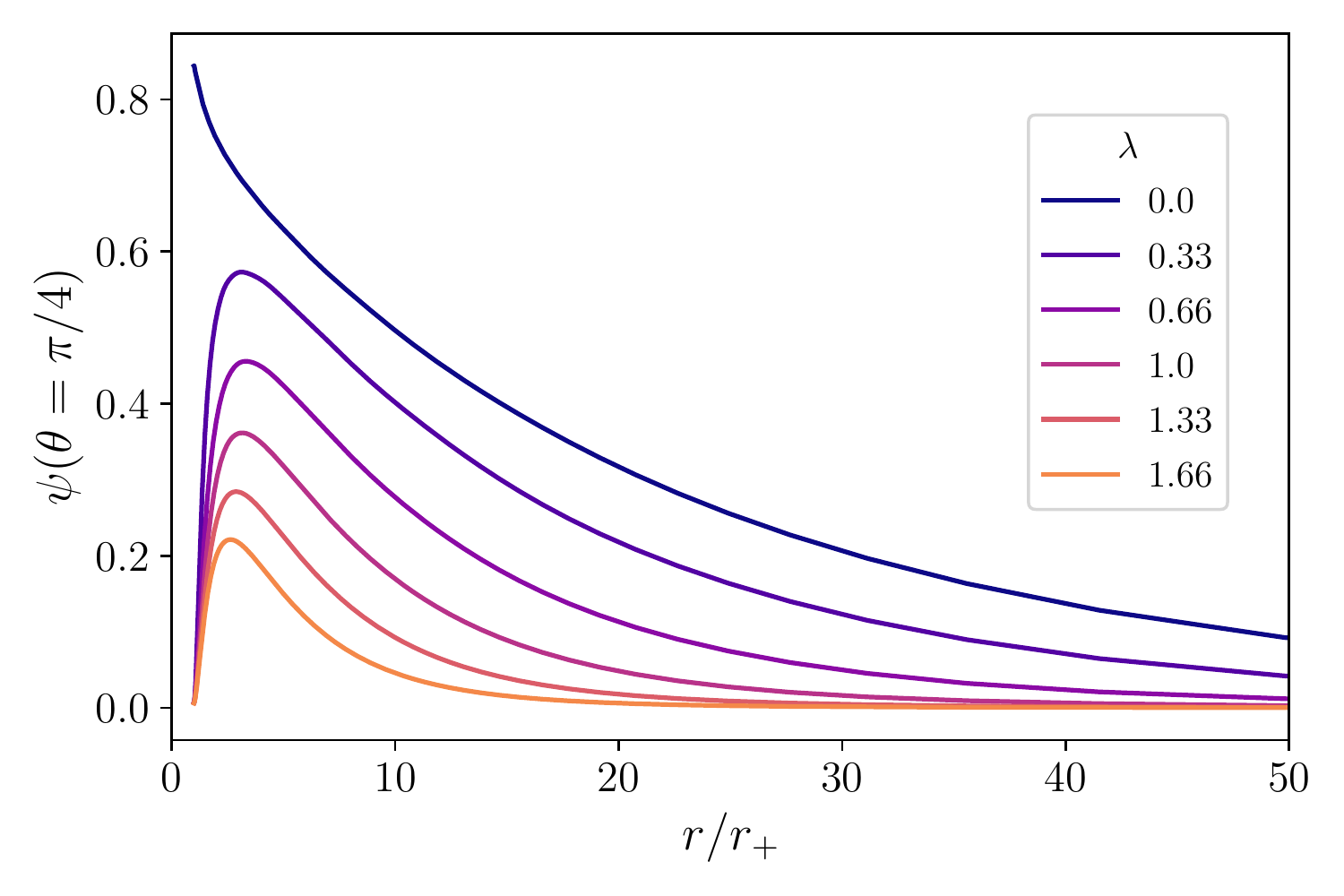}
\caption{A closer look on the axion cloud gap near the wormhole throat. Here we show slices of $\psi$ for constant $\theta = \pi/4$, with parameters $a = 0.99, \mu^2 = 0^+$.
Increasing of $\lambda$ decreases the magnitude of the axions and cuts off its tail.}
\label{fig:kw_slices}
\end{figure}

The radial behaviour of the axionic field can be seen more precisely in Fig. \ref{fig:kw_slices}.
We present there a slice of $\psi$ as a function of $r$ in throat radius units, for constant $\theta=\pi/4$ and few different values of the $\lambda$ parameter.
In contrast, we also plot the behaviour of axions in Kerr black hole metric (that is $\lambda = 0$).

What we can see is the increasing $\lambda$ consequently extinguishes the axionic hair.
In the foreground a structural change in the field profile is visible as we compare it to the black hole scenario.
An axionic field over a black hole has a maximum value on the event horizon. 
The opposite is true for a wormhole, on the throat the field vanishes, then grows to its maximum and fades away with the radius.
Then, the bigger is $\lambda$ the smaller are the maxima and overall magnitude of the axionic hair.

\begin{figure}
\centering
\includegraphics[width=0.8\textwidth]{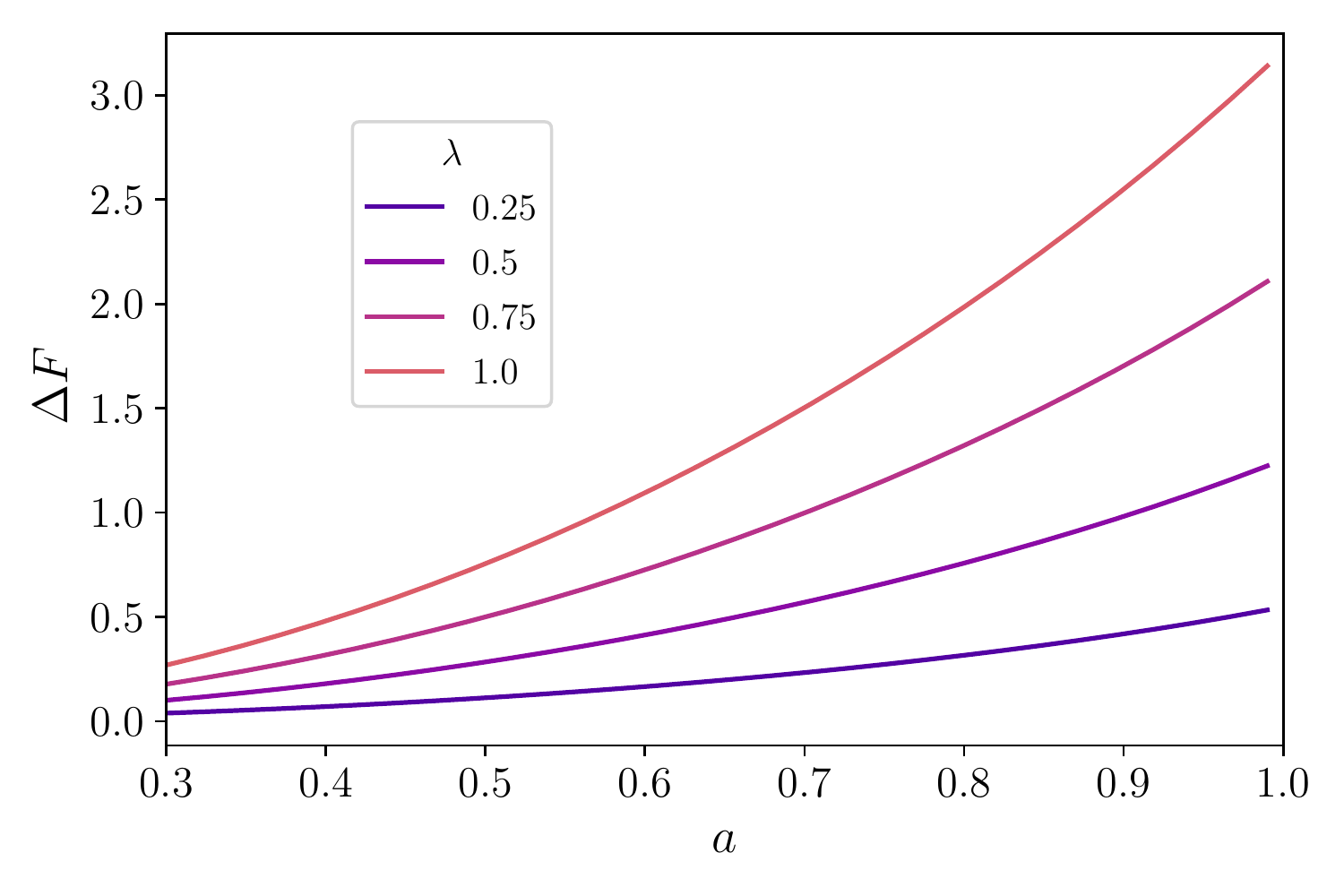}
\caption{Free energy differences as a function of angular momentum (with $\lambda = 0$ as the ground curve) for different values of $\lambda$.  With axion mass $\mu^2 = 0^+$ we see that the cloud
thermodynamical favourability decreases with growth of both angular momentum and $\lambda$ parameter.}
\label{fig:kw_fe}
\end{figure}

In the next step we investigate the free energy of the obtained axion cloud configurations.
It is interesting to see how the parameters describing the spacetime geometry around the wormhole influence the thermodynamics of the axion clouds.
Due to the previously mentioned difficulties in computing the free energy in these metrics, we rather talk about energy differences, than the exact values.
In Fig. \ref{fig:kw_fe} we present the differences of the free energy versus angular momentum $a$, with respect to the $\lambda=0$ level, which constitutes a plain Kerr black hole.
It is clearly visible that the larger value of the distortion parameter $\la$ one takes into account, the higher value of the free energy of the cloud we achieve.
It turns out, that the more the gravitational background deviates from from the black hole metric, the less thermodynamically desirable axion clouds are.
This effect works together with the diminishing magnitude of the field on the previously discussed Fig. \ref{fig:kw_slices}.
Additionally the increasing angular momentum of the wormhole also increases the free energy difference. 
This means that for \textit{extreme} Kerr-like wormhole axion hair is the least favourable.

\subsection{Teo rotating wormhole}
Teo class wormhole has a different set of parameters and it does not simply transform into a black hole solution, just like a Kerr-like wormhole does.
Here the throat radius is independent of the other parameters and is imposed manually in $g_{tt}$ and $g_{rr}$.
With the particular choice of functions \eqref{eq:twh_metric_fun}, we can only steer with the shape of metric components via $\gamma$ parameter.
Therefore, let us consider values of $\gamma$ in the interval $(-1, 1]$,
where for $\gamma = -1$ the function $g_{rr}$ is singular, so we can only approach this value.

\red{As it was mentioned, the Kerr-like wormhole can be reduced to a black hole solution by setting $\lambda = 0$. 
Teo solution does not share this feature, but is a well-known wormhole metric, just like its non-rotating counterpart the archetypical Morris-Thorne wormhole.
While it can not serve as a testing field for differences between axion clouds around wormholes and black holes, one can treat it as a benchmark for behaviours of the axion hair in another wormhole environment.
Using this background might help us to see if the obtained axion solutions share similar features.
}

\begin{figure}[h]
\centering
\subfloat[$\mu^2 \rightarrow 0^+, \quad \gamma = -0.99$]{
\includegraphics[width=0.45 \textwidth]{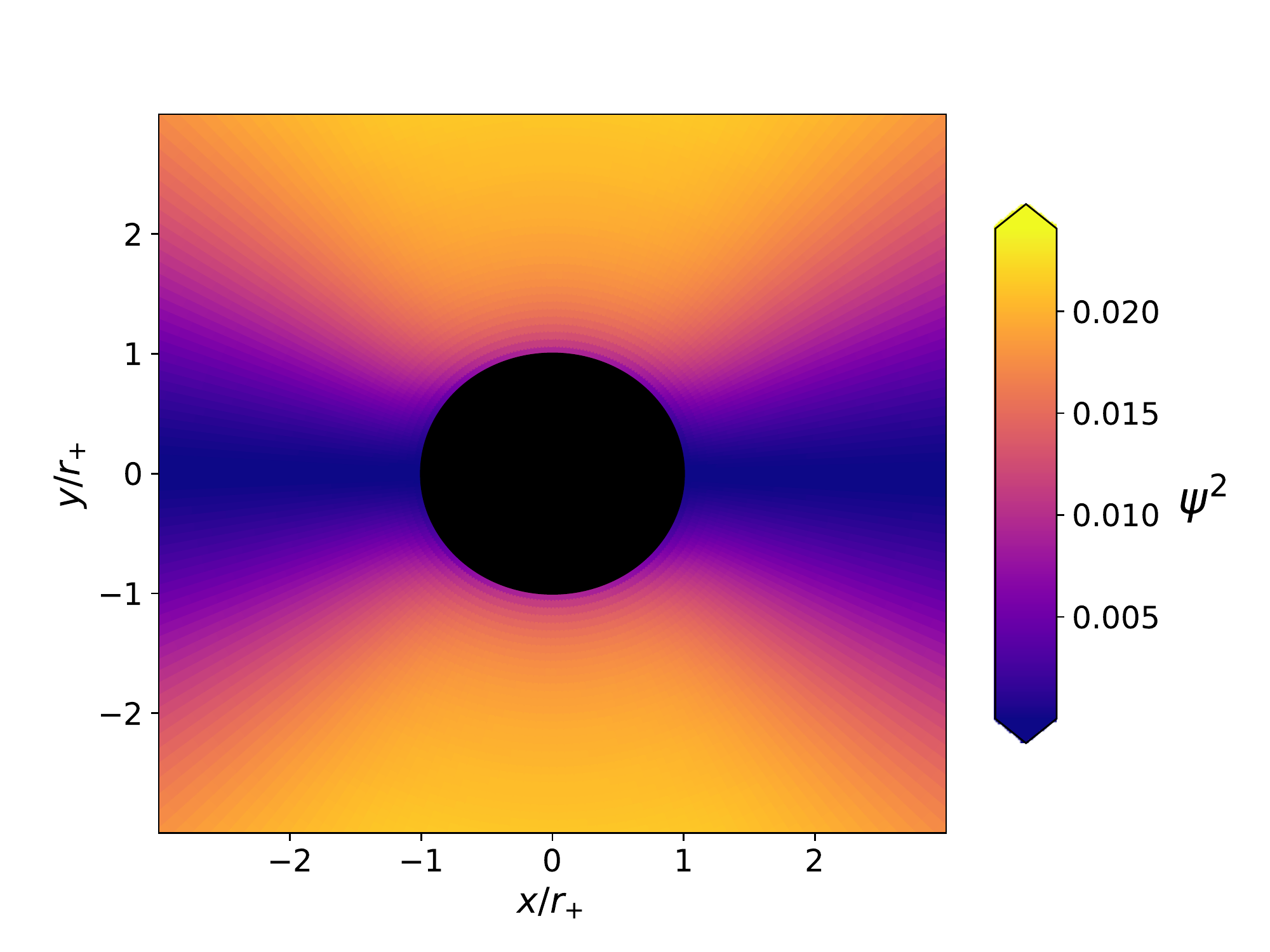}
}
\qquad
\subfloat[$\mu^2 = 0.1, \quad \gamma = -0.99$]{
\includegraphics[width=0.45 \textwidth]{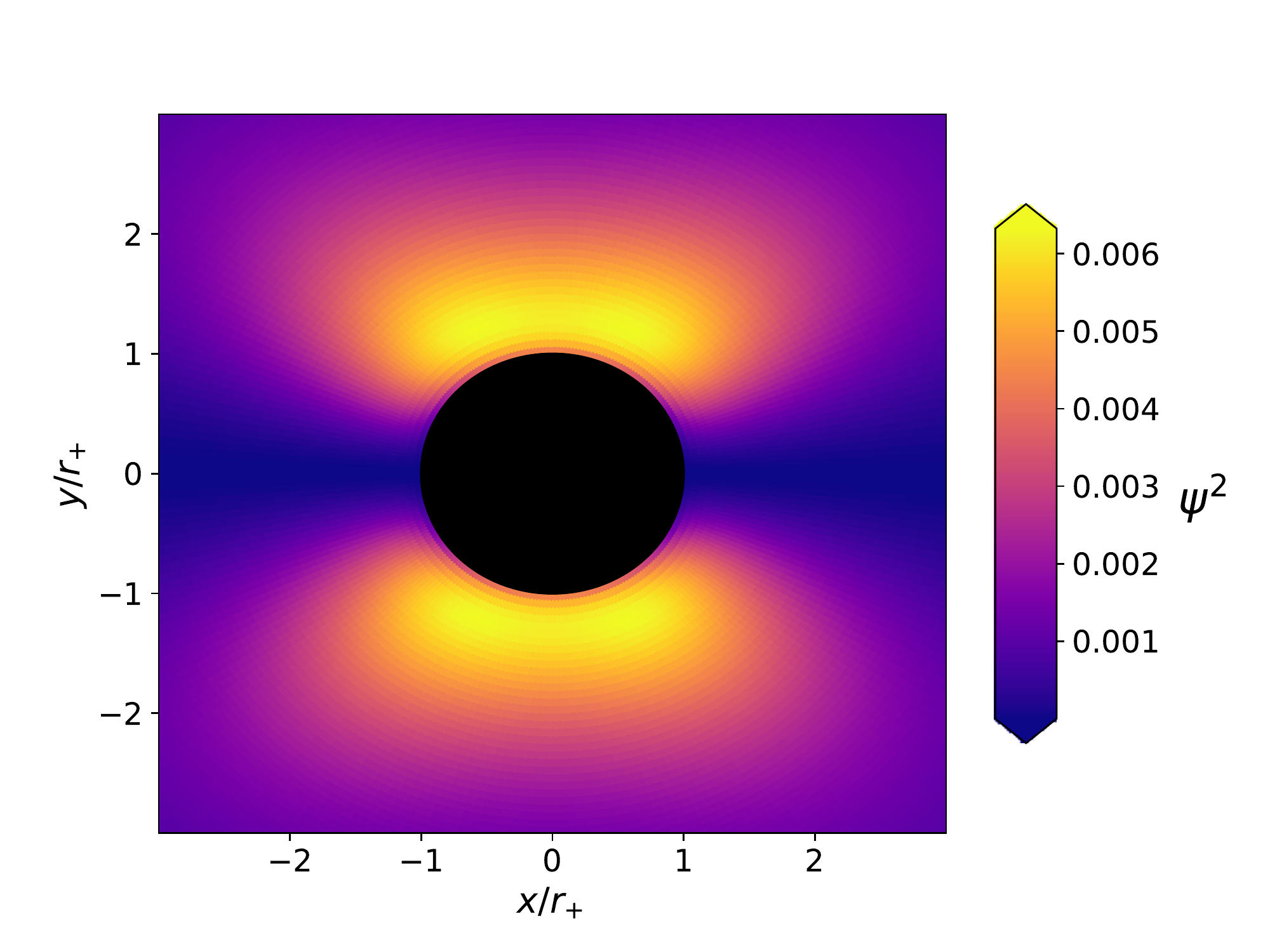}
}

\vspace{0.5cm}
\subfloat[$\mu^2 = 0.5, \quad  \gamma = -0.99$]{
\includegraphics[width=0.45 \textwidth]{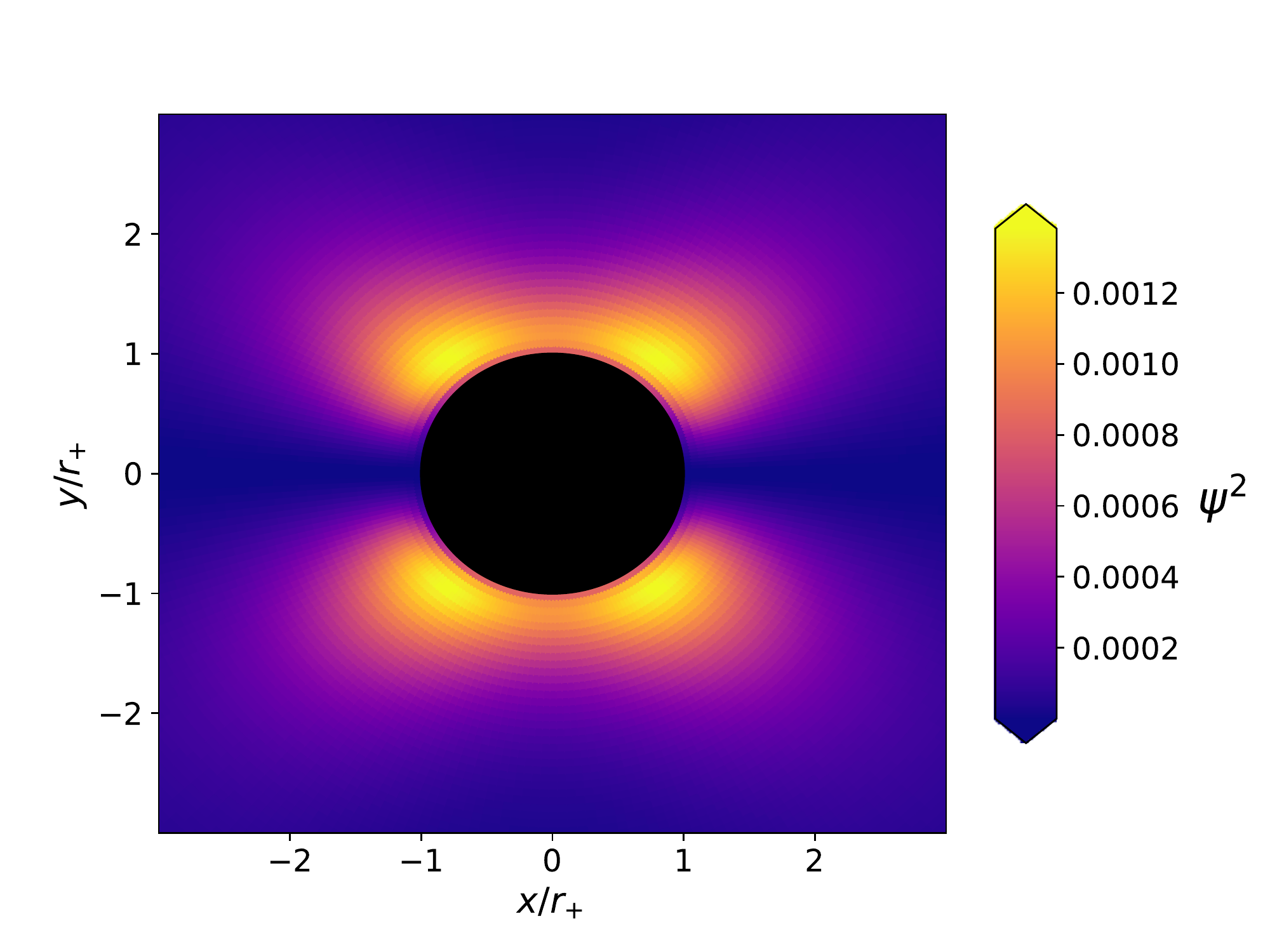}
}
\qquad
\subfloat[$\mu^2 = 1, \quad \gamma = -0.99$]{
\includegraphics[width=0.45 \textwidth]{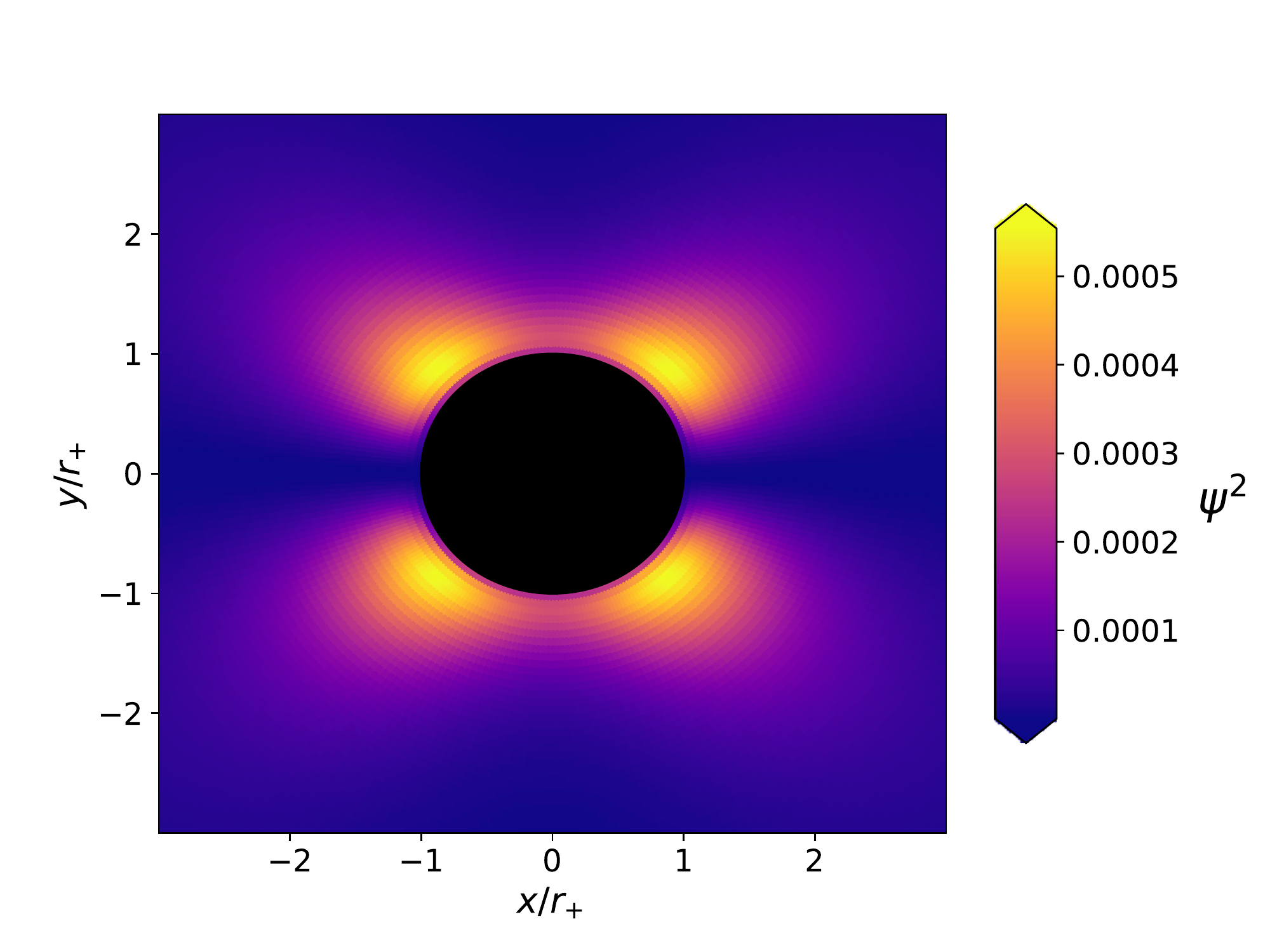}
}
\caption{Axion cloud distribution around the magnetized Teo type wormhole.  The negative value of $\gamma$ resembles the results for Kerr-like wormhole, with 
large $\lambda$ value, i.e., the decreasing of the field magnitude.
Each panel depicts the field distribution for different axionlike masses.
Once again a big mass results with a localized field, concentrated closely to the throat with depleted polar region. For bigger values of $\gamma$ we observe similar influence of $\mu$ on the spatial distribution.}
\label{fig_teo_maps}
\end{figure}

In Fig. \ref{fig_teo_maps} one can see the distribution of the axionic cloud around a Teo wormhole for different axionlike field masses.
In the panel (a) we have an ultralight field, then it takes values $0.1$, $0.5$ and $1.0$ for (b), (c) and (d) respectively.
In all panels we use $\gamma = -0.99$, which gives us tiny value of the axionic field (see the colorbars).

The angular distribution has similar features to the Kerr-like metric. 
For ultralight axions the field is localized in the majority of the wormhole surroundings.
As the mass increases the hair tightens spatially and disappears from the polar regions.
For the large mass case the axionic clouds are drifting toward the equator with the polar caps left almost empty. Moreover the radial reach is very short - around one throat radius.

One can clearly notice that for the negative gamma value the analogous effect to the Kerr-like scenario is observed.
The axionic cloud is also pushed away from the throat surface - in its vicinity the field acquires small values, reaches the maximum and the descends monotonically to zero.
However in this gravitational background this effect is not as dramatic as in case of Kerr-like wormhole.
The weakening of the axionic field in the vicinity of the throat is easily visible in the distributions, although it is not that large.
Increasing $\gamma$ up to zero and beyond causes the rise of $\psi$ field. It grants bigger values, but the spatial qualitative characteristic remains intact.

\begin{figure}[h]
\centering
\includegraphics[width=0.7 \textwidth]{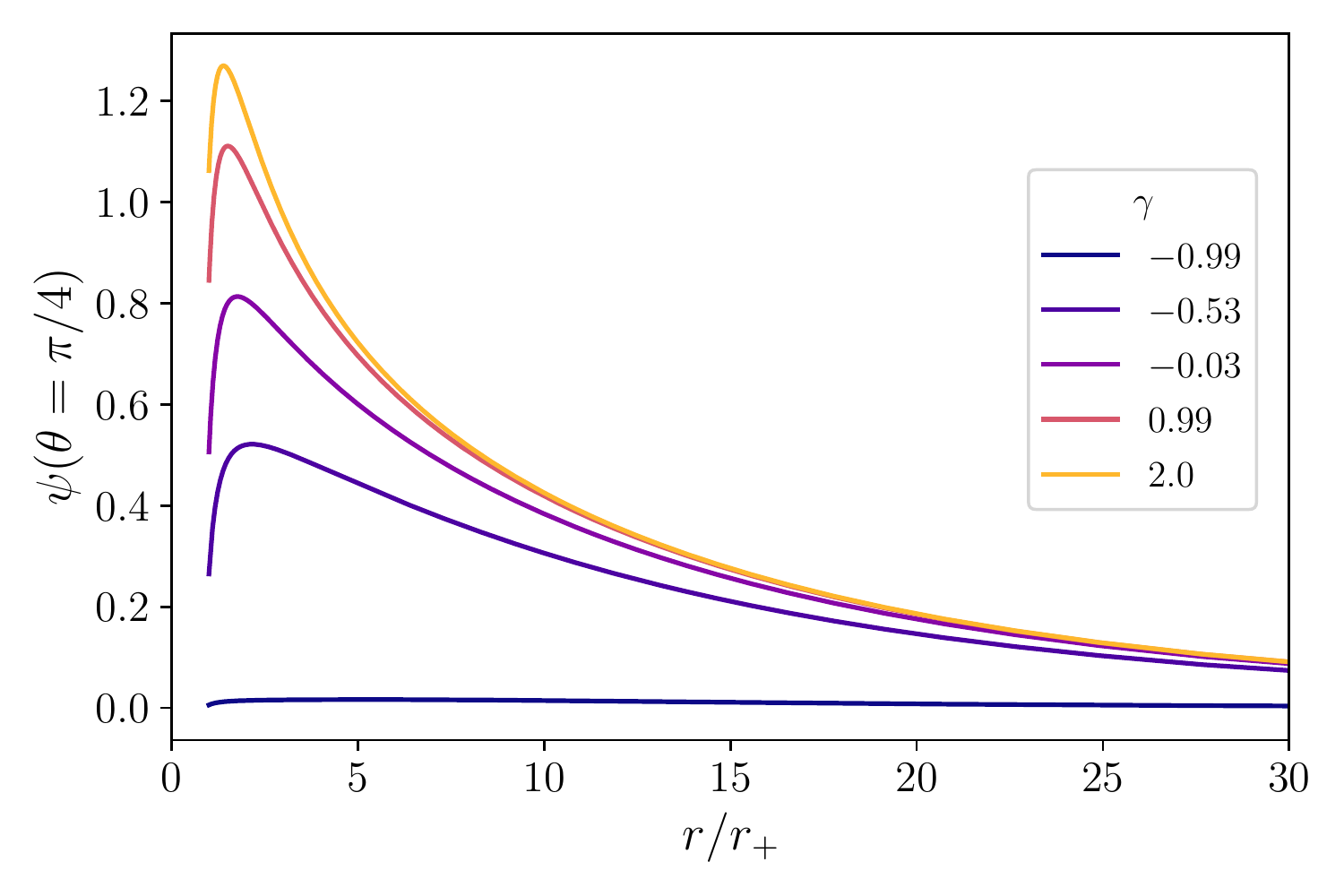}
\caption{Radial slices of the axion field for constant angle $\theta = \pi/4$. 
The growth of $\gamma$ increases the maximum value of the field. 
However in this metric the significant growth of the blank space near the throat is not present.
Also the growth of $\gamma$ does not seem to greatly affect the tail of the field, away from the throat, which is different from the Kerr-like wormhole results.}
\label{fig_teo_slice}
\end{figure}

The next figure brings a closer look on the field drop near the throat.
In Fig.  \ref{fig_teo_slice} we present radial slices of the field distribution with $\theta = \pi/4$. In this particular figure we depict the behaviour for ultralight field, however a similar tendencies are shown by more massive fields.
First of all, we observe a significant amplification of the field with the growth of $\gamma$. 
The field does not acquire new features however, but it seems that the curves follow some kind of scaling related to gamma.
Additionally the profiles resemble the results obtained for Kerr-like wormholes. As we have previously mentioned, Teo class wormhole cannot be simply transformed into a black hole by a simple choice of parameter value.
However the features like a drop near the throat surface, then maximum and monotonic fall show that these might be more general wormhole related behaviours of axionic hair.

\begin{figure}[h]
\centering
\includegraphics[width=0.7 \textwidth]{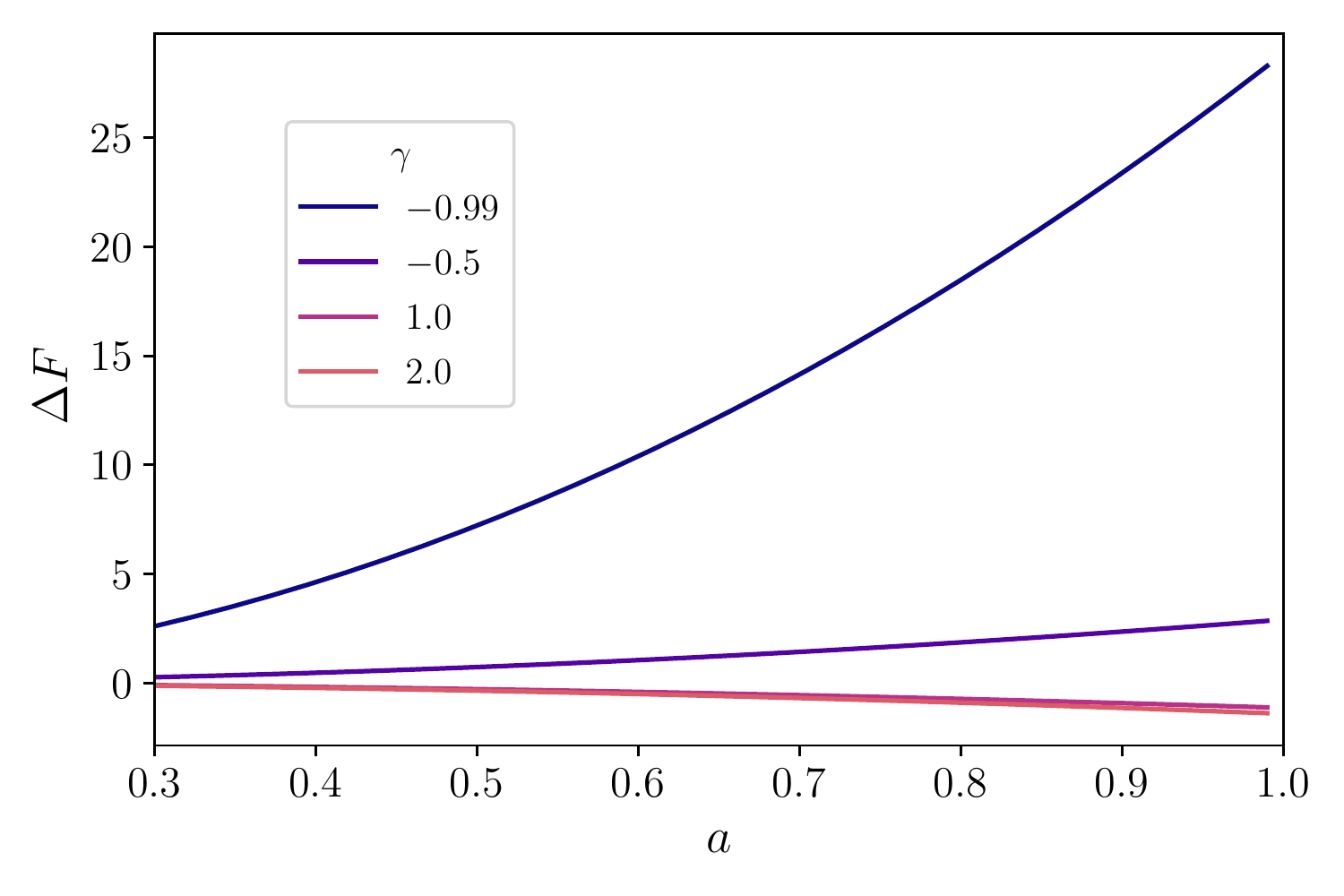}
\caption{Free energy differences vs.  angular momentum for different values of $\gamma$. The curve of $\gamma = 0$ is the reference level.  We see that free energy increases for negative values of gamma and slightly drops for positive ones, as angular momentum grows.}
\label{fig_teo_fe}
\end{figure}

If we consider the free energy, it appears that the axion clouds for the background with negative $\gamma$, are definitely less thermodynamically favourable.
In Fig. \ref{fig_teo_fe} we plot the free energy difference as a function of angular momentum for several values of the gamma parameter.
We use the curve with $\gamma = 0$ as a baseline for calculating energy differences.

Free energy difference curves for $\gamma < 0$ are positive, especially the $\gamma = -0.99$ curve reaches relatively big values.
Therefore thermodynamically speaking wormholes with $\gamma$ close to $-1$ have least chances to hold axionic hair.
With the growth of the parameter, the free energy of the cloud decreases, which makes the axions more thermodynamically favourable.
However this fall is rather moderate comparing to the rise of the top curve.

In both cases the increase of angular momentum amplifies the tendencies of the curves.
Curves for negative gamma grow, while the positive fall.
A consequent growth of $\gamma$ parameter leads the hair to some limit characteristics which can be seen in Fig. \ref{fig_teo_slice} and Fig.  \ref{fig_teo_fe}.

Finally let us conduct qualitative comparison of the axionic clouds in considered metrics.
The solutions have undoubtedly similar features, especially when one takes a look on the $\psi$ slices.
We also observe the separation of the cloud from the surface of the throat in both cases.
This allows us to notice some general wormhole related phenomena, which are not present around the black holes \cite{kic21}.
Naturally we cannot speak in a fully general manner, as we only considered here merely two distinct gravitational backgrounds, which on the other hand should be treated as toy models.

\section{Conclusion}
In our paper, we have considered the problem of the distribution of axionlike particles, being regarded as {\it dark matter} sector model, around 
the toy models of rotating wormholes. We have investigated the Kerr-like wormhole line element with the distortion parameter $\la$ and
Teo model of a rotating axisymmetric wormhole. The models under inspection were characterized by the mass, angular momentum, distortion parameter
(for Kerr-like wormhole) and the shape parameter $\ga$ (for the Teo model). We numerically solve the equations of motion for the underlying cases, using spectral methods.

Among all we have found that the axion clouds are pushed forward from the wormhole throat, especially for the case of large value of the rotation
parameter $a$. The voids in the vicinity of the wormhole throat appear for the larger value of the distortion parameter $\la$. This phenomenon distinguishes the studied
system from the previously elaborated black holes \cite{kic21}. On the other hand, for the larger $\la$, one achieves higher value of the free energy, and therefore
this solution is less thermodynamically favoured.

As far as the Teo class of rotating wormholes is concerned, we have for the negative value of $\ga$ the analogous effect as in the latter case is obtained. However
for the positive value, the behavior of axionic clouds resembles features of the {\it dark matter} clouds around Kerr black hole in a uniform magnetic field.
The solution with negative $\ga$ is not thermodynamically favourable, as it has been revealed in free energy studies. However when $\ga$ increases, the free energy
of the axionic cloud decreases.

We have found that the behavior of the axionic clouds significantly differs from the black hole scenario, which we discussed in our previous work \cite{kic21}.
This fact will account for the possible guidance, enabling one to distinguish between these two classes of compact objects.
Nevertheless, the search for astronomically observable criteria require far more complex approach. A more realistic dynamical gravitational model is needed, 
when the time dependence of the studied fields is taken into account, as well as, the direct mathematical proofs of the stabilities of rotating wormhole spacetimes ought to be found.
These subjects impose a real mathematical challenge and require also solid numerical relativity machinery. 
These problems shall be investigated elsewhere.



\begin{thebibliography}{99}

%
\def\cmp#1#2#3#4{\emph{#4}, \emph{ Commun. Math. Phys.} {\bf #1} (#3) #2}
\def\lmp#1#2#3#4{\emph{#4}, \emph{ Lett. Math. Phys.} {\bf #1} (#3) #2}
\def\hpa#1#2#3#4{\emph{#4}, \emph{ Hell. Phys. Acta} {\bf #1} (#3) #2}
\def\grg#1#2#3#4{\emph{#4}, \emph{ Gen. Rel. Grav.} {\bf #1} (#3) #2}
\def\pr#1#2#3#4{\emph{#4}, \emph{ Phys. Rev.} {\bf #1} (#3) #2}
\def\prl#1#2#3#4{\emph{#4}, \emph{ Phys. Rev. Lett.} {\bf #1} (#3) #2}
\def\prd#1#2#3#4{\emph{#4}, \emph{ Phys. Rev. D} {\bf #1} (#3) #2}
\def\prb#1#2#3#4{\emph{#4}, \emph{ Phys. Rev. B} {\bf #1} (#3) #2}
\def\prx#1#2#3#4{\emph{#4}, \emph{ Phys. Rev. X} {\bf #1} (#3) #2}
\def\pl#1#2#3#4{\emph{#4}, \emph{ Phys. Lett.} {\bf #1} (#3) #2}
\def\pla#1#2#3#4{\emph{#4}, \emph{ Phys. Lett. A} {\bf #1} (#3) #2}
\def\plb#1#2#3#4{\emph{#4}, \emph{ Phys. Lett. B} {\bf #1} (#3) #2}
\def\prep#1#2#3#4{\emph{#4}, \emph{ Phys. Reports} {\bf #1} (#3) #2}
\def\phys#1#2#3#4{\emph{#4}, \emph{ Physica} {\bf #1} (#3) #2}
\def\jcp#1#2#3#4{\emph{#4}, \emph{ J. Comput. Phys.} {\bf #1} (#3) #2}
\def\jmp#1#2#3#4{\emph{#4}, \emph{ J. Math. Phys.} {\bf #1} (#3) #2}
\def\jpm#1#2#3#4{\emph{#4}, \emph{ J. Phys. A: Math. Gen.} {\bf #1} (#3) #2}
\def\cpr#1#2#3#4{\emph{#4}, \emph{ Computer Phys. Rept.} {\bf #1} (#3) #2}
\def\cqg#1#2#3#4{\emph{#4}, \emph{ Class. Quant. Grav.} {\bf #1} (#3) #2}
\def\cma#1#2#3#4{\emph{#4}, \emph{ Computers Math. Applic.} {\bf #1} (#3) #2}
\def\mc#1#2#3#4{\emph{#4}, \emph{ Math. Compt.} {\bf #1} (#3) #2}
\def\apj#1#2#3#4{\emph{#4}, \emph{ Astrophys. J.} {\bf #1} (#3) #2}
\def\apjs#1#2#3#4{\emph{#4}, \emph{ Astrophys. J. Suppl.} {\bf #1} (#3) #2}
\def\apjl#1#2#3#4{\emph{#4}, \emph{ Astrophys. J. Lett.} {\bf #1} (#3) #2}
\def\acta#1#2#3#4{\emph{#4}, \emph{ Acta Astronomica} {\bf #1} (#3) #2}
\def\apl#1#2#3#4{\emph{#4}, \emph{ Ann. Physik. (Leipzig)} {\bf #1} (#3) #2}
\def\amjp#1#2#3#4{\emph{#4}, \emph{Am. J. Phys.} {\bf #1} (#3) #2}
\def\anp#1#2#3#4{\emph{#4}, \emph{ Ann. Phys.} {\bf #1} (#3) #2}
\def\sa#1#2#3#4{\emph{#4}, \emph{ Sov. Astro.} {\bf #1} (#3) #2}
\def\sia#1#2#3#4{\emph{#4}, \emph{ SIAM J. Sci. Statist. Comput.} {\bf #1} (#3) #2}
\def\aa#1#2#3#4{\emph{#4}, \emph{ Astron. Astrophys.} {\bf #1} (#3) #2}
\def\mnras#1#2#3#4{\emph{#4}, \emph{ Mon. Not. R. Astr. Soc.} {\bf #1} (#3) #2}
\def\npb#1#2#3#4{\emph{#4}, \emph{ Nucl. Phys. B} {\bf #1} (#3) #2}
\def\npa#1#2#3#4{\emph{#4}, \emph{ Nucl. Phys. A} {\bf #1} (#3) #2}

\def\prsla#1#2#3#4{\emph{#4}, \emph{ Proc. R. Soc. London, Ser. A} {\bf #1} (#3) #2}
\def\jhep#1#2#3#4{\emph{#4}, \emph{ JHEP} {\bf #1} (#2) #3}
\def\jcap#1#2#3#4{\emph{#4}, \emph{ JCAP} {\bf #1} (#2) #3}

\def\nuca#1#2#3#4{\emph{#4}, \emph{ Nuovo Cimento A } {\bf #1} (#3) #2}
\def\nucb#1#2#3#4{\emph{#4}, \emph{ Nuovo Cimento B } {\bf #1} (#3) #2}
\def\ijmp#1#2#3#4{\emph{#4}, \emph{ Int. J. Mod. Phys. D} {\bf #1} (#3) #2}
\def\atmp#1#2#3#4{\emph{#4}, \emph{ Adv. Theor. Math. Phys.} {\bf #1} (#3) #2}
\def\ptps#1#2#3#4{\emph{#4}, \emph{ Prog. Theor. Phys. Suppl.} {\bf #1} (#3) #2}
\def\ptp#1#2#3#4{\emph{#4}, \emph{ Prog. Theor. Phys.} {\bf #1} (#3) #2}
\def\lmp#1#2#3#4{\emph{#4}, \emph{ Lett. Math. Phys.} {\bf #1} (#3) #2}
\def\cpam#1#2#3#4{\emph{#4}, \emph{ Comm. Pure Appl. Math.}  {\bf #1} (#3) #2}
\def\adv#1#2#3#4{\emph{#4}, \emph{ Adv. Phys.}  {\bf #1} (#3) #2}
\def\zh#1#2#3#4{\emph{#4}, \emph{ Zh. Eksp. Teor. Fiz.}  {\bf #1} (#3) #2}
\def\mplb#1#2#3#4{\emph{#4}, \emph{ Mod. Phys. Lett. B} {\bf #1} (#3) #2}
\def\jams#1#2#3#4{\emph{#4}, \emph{ J. Austral. Math. Soc. B} {\bf #1} (#3) #2}
\def\appa#1#2#3#4{\emph{#4}, \emph{ Acta Phys. Polonica A} {\bf #1} (#3) #2}
\def\appb#1#2#3#4{\emph{#4}, \emph{ Acta Phys. Polonica B} {\bf #1} (#3) #2}
\def\nat#1#2#3#4{\emph{#4}, \emph{Nature} {\bf #1} (#3) #2}
\def\natcom#1#2#3#4{\emph{#4}, \emph{Nature Commun.} {\bf #1} (#3) #2}
\def\natphys#1#2#3#4{\emph{#4}, \emph{Nature Physics} {\bf #1} (#3) #2}
\def\natmat#1#2#3#4{\emph{#4}, \emph{Nature Mat.} {\bf #1} (#3) #2}


\def\science#1#2#3#4{\emph{#4}, \emph{Science} {\bf #1} (#3) #2}
\def\sciadv#1#2#3#4{\emph{#4}, \emph{Sci. Adv.} {\bf #1} (#3) #2}

\def\arcmp#1#2#3#4{\emph{#4}, \emph{Annual Rev. of Cond. Matter Physics} {\bf #1} (#3) #2}
\def\zphys#1#2#3#4{\emph{#4}, \emph{Z. Phys.} {\bf #1}, (#3) #2}
\def\ncs#1#2#3#4{\emph{#4}, \emph{Nuovo Cimento Suppl.} {\bf #1} (#3) #2}
\def\physb#1#2#3#4{\emph{#4}, \emph{Physica B} {\bf #1}, (#3) #2}
\def\jpcm#1#2#3#4{\emph{#4}, \emph{J. Phys.: Condens. Matter } {\bf #1} (#3) #2}
\def\pnas#1#2#3#4{\emph{#4}, \emph{Proc. Nat. Academy Sciences} {\bf #1} (#3) #2}
\def\sssr#1#2#3#4{\emph{#4}, \emph{Izv. Akad Nauk SSSR, ser. fiz.} {\bf #1} (#3) #2}
\def\jpg#1#2#3#4{\emph{#4}, \emph{ J. Phys. G} {\bf #1} (#3) #2}
\def\chinpb#1#2#3#4{\emph{#4}, \emph{Chin. Phys. B} {\bf #1} (#3) #2}
\def\njp#1#2#3#4{\emph{#4}, \emph{ New J. Phys.} {\bf #1} (#3) #2}
\def\frontphys#1#2#3#4{\emph{#4}, \emph{ Front. Phys.} {\bf #1} (#3) #2}
\def\epl#1#2#3#4{\emph{#4}, \emph{ EPL} {\bf #1} (#3) #2}
\def\rmp#1#2#3#4{\emph{#4}, \emph{ Rev. Mod. Phys.} {\bf #1} (#3) #2}
\def\aphss#1#2#3#4{\emph{#4}, \emph{ Astrophys. Space Sci.} {\bf #1} #2 (#3)}

\def\hepph#1#2{{ hep-ph }{#1} (#2)}
\def\arxiv#1#2#3{\emph{#3},{ arXiv }{#1} (#2)}
\def\hepth#1#2{{ hep-th }{#1} (#2)}
\def\grqc#1#2{{ gr-qc }{#1} (#2)}
\def\ibid#1#2#3#4{\emph{#4}, {\it ibid.} {\bf #1} (#3) #2}
\def\conphy#1#2#3#4{\emph{#4}, \emph{Contemporary Physics} {\bf #1}, (#3) #2}
\def\ppnp#1#2#3#4{\emph{#4}, \emph{ Prog. Part. Nucl. Phys} {\bf #1} (#3) #2}
\def\arnps#1#2#3#4{\emph{#4}, \emph{ Annu. Rev. Nucl. Part. Sci.} {\bf #1} (#3) #2}
\def\pz#1#2#3#4{\emph{#4}, \emph{ Phys. Z.} {\bf #1} (#3) #2}





\bibitem{fla16}
L. Flamm, \pz{17}{448}{1916}{Beitr\"age zur Einsteinischen Gravitationstheorie}.
\bibitem{ein35}
A. Einstein and N. Rosen, \pr{48}{73}{1935}{The particle problem in the general theory of relativity}.
\bibitem{whe55}
J. A. Wheeler, \pr{97}{511}{1955}{Geons}.
\bibitem{haw88}
S. W. Hawking, \prd{37}{904}{1988}{Wormholes in space-time}.
\bibitem{mor88}
M. S. Morris and K. Thorne, \amjp{56}{395}{1988}{Wormholes in space-time and their use for interstellar travel: A tool for teaching general relativity}.
\bibitem{ell73}
H. G. Ellis, \jmp{14}{104}{1973}{Ether flow through a drainhole: A particle model in general relativity}.
\bibitem{bro73}
K. Bronikov, \appb{4}{251}{1973}{Scalar-Tensor Theory and Scalar Charge}.
\bibitem{ell79}
H. G. Ellis, \grg{10}{105}{1979}{The evolving, flowless drainhole: A nongravitating-particle model in general relativity theory}.

\bibitem{kan11}
P. Kanti, B. Kleinhaus, and J. Kunz, \prl{107}{271101}{2011}{Wormholes in dilatonic Einstein-Gauss-Bonnet Theory}.
\bibitem{har13}
T. Harko, F. S. N.Lobo, M. K. Mak, and S. V. Sushkov, \prd{87}{067504}{2013}{Modified-gravity wormholes without exotic matter}.
\bibitem{gib16}
G. W. Gibbons and M. S. Volkov, \plb{760}{324}{2016}{Ring wormholes via duality rotations}.
\bibitem{gib17}
G. W. Gibbons and M. S. Volkov, \jcap{05}{039}{2017}{Weyl metrics and wormholes}.
\bibitem{gou18}
P. Goulart, \cqg{35}{025012}{2018}{Phantom wormholes in Einstein-Maxwell-dilaton theory}.

\bibitem{teo98}
E. Teo, \prd{58}{024014}{1998}{Rotating tranversable wormholes}.
\bibitem{cle84}
G. Clement, \grg{16}{131}{1984}{A class of wormhole solutions to higher-dimensional general relativity}.
\bibitem{bro13}
K. A. Bronikov, V. G. Krechet, and J. P. S. Lemos, \prd{87}{084060}{2013}{Rotating cylindrical wormholes}.
\bibitem{kas08}
P. E. Kashargin and S. V. Sushkov, \prd{78}{064071}{2008}{Slowly rotating scalar field wormholes: The second order approximation}.
\bibitem{kle14}
B. Kleinhaus and J. Kunz, \prd{90}{121503}{2014}{Rotating Ellis wormholes in four dimensions}.
\bibitem{che16}
X. Y. Chew, B. Kleinhaus, and J. Kunz, \prd{94}{104031}{2016}{Geometry of spinning Ellis wormholes}.
\bibitem{vol21}
M. S. Volkov, \prd{104}{124064}{2021}{Stationary generalization for Bronnikov-Ellis wormhole and for the vacuum ring wormhole}.


\bibitem{mat06}
T. Matos and D. Nunez, \cqg{23}{4485}{2006}{Rotating scalar field wormhole}.
\bibitem{rub89}
P. J. Ruback, \cqg{6}{L21}{1989}{A uniqueness theorem for wormholes in quantum gravity}.
\bibitem{yaz17}
S. Yazadjiev, \prd{96}{044045}{2017}{Uniqueness theorem for static wormholes in Einstein phantom scalar field theory}.
\bibitem{laz17}
B. Lazov, P. Nedkova, and S. Yazadjiev, \plb{778}{408}{2018}{Uniqueness theorem for static phantom wormholes in Einstein-Maxwell-dilaton theory}.
\bibitem{rog18}
M. Rogatko, \prd{97}{024001}{2018}{Uniqueness of higher-dimensional phantom field wormholes}.
\bibitem{rog18a}
M. Rogatko, \prd{97}{064023}{2018}{Uniqueness of higher-dimensional Einstein-Maxwell-phantom dilaton field wormholes}.





\bibitem{worm}
M. Visser, {\it Lorentzian Wormholes}, New York, American Institute of Physics 1995,~
F. S. N. Lobo ed., {\it Wormholes, warp drives and energy conditions}, New York, Springer Fundamentals in Theor. Phys., 2017.







\bibitem{dam07}
T. Damour and S. N. Solodukhin, \prd{76}{024016}{2007}{Wormholes as black holes foils}.
\bibitem{bue18}
P. Bueno, P. A. Cano, F. Goelen, T. Hertog, and B. Vercnocke, \prd{97}{024040}{2018}{Echoes of Kerr-like wormholes}.
\bibitem{ami19}
M. Amir, K. Jusufi, A. Banerjee, and S. Hansraj, \cqg{36}{215007}{2019}{Shadow images of Kerr-like wormholes}.
\bibitem{kas21}
S. Kasuya and M. Kobayashi, \prd{103}{104050}{2021}{Throat effects on shadows of Kerr-like wormholes}.



\bibitem{kic21}
B. Kiczek and  M. Rogatko, \prd{103}{124021}{2021}{Axionlike dark matter clouds around rotating black holes}.

\bibitem{pre83}
J. Preskill, M. B. Wise, and F. Wilczek, \plb{120}{127}{1983}{Cosmology of the Invisible Axion}.
\bibitem{abb83}
 L. F. Abbott and P. Sikivie, \plb{120}{133}{1983}{ Cosmological Bound on the Invisible Axion}.
\bibitem{din83} 
M. Dine and W. Fischler, \plb{120}{137}{1983}{The Not So Harmless Axion}.
\bibitem{pec77}
R. D. Peccei and H. R. Quinn, \prl{38}{1440}{1977}{CP Conservation in the Presence of Pseudoparticles}.
\bibitem{wei78}
S. Weinberg, \prl{40}{223}{1978}{A New Light Boson?}.
\bibitem{wil78}
F. Wilczek, \prl{40}{279}{1978}{Problem of Strong $P$ and $T$ Invariance in the Presence of Instantons}.

\bibitem{pen15}
J. M. Pendlebury et al., \prd{92}{092003}{2015}{Revised experimental upper limit on the electric dipole moment of the neutron}.
\bibitem{svr06}
P. Svrcek and E. Witten, \jhep{06}{2006}{051}{Axions in string theory}.




\bibitem{pla18}
A. D. Plascencia and A. Urbano, \jcap{04}{2018}{059}{Black hole supperadiance and polarization-dependent bending of light}.
\bibitem{gao20}
X. Bi, Y. Gao, J. Guo, N. Houston, T. Li, F. Xu, and X. Zhang, \arxiv{2002.01796}{2020}{Axion and dark photon limits from Crab Nebula high energy gamma-rays}.
\bibitem{ike19}
T. Ikeda, R. Brito, and V. Cardoso, \prl{122}{081101}{2019}{Blasts of light from axions}.
\bibitem{bos19}
M. Boskovic, R. Brito, V. Cardoso, T. Ikeda, and H. Witek, \prd{99}{035006}{2019}{Axionic instabilities and new black hole solutions}.
\bibitem{car18}
V. Cardoso, O. J. C. Dias, G. S. Harnett, M. Middleton, P. Pani, and J. E. Santos, 
\jcap{03}{2018}{043}{Constraining the mass of dark photons and axion-like particles through black-hole superradiance}.

\bibitem{gar18}
B. Garbrecht and J. I. Mc Donald, \jcap{07}{2018}{044}{Axion configurations around pulsars}.
\bibitem{dar20a}
J. Darling, \apjl{900}{L28}{2020}{New limits on axionic dark matter using the magnetar PSR J1745-2900}.
\bibitem{dar20b}
J. Darling, \prl{125}{121103}{2020}{Search for axion dark matter using the magnetar PSR J1745-2900}.
\bibitem{gra15}
P. W. Graham, I. G. Irastorza, S. K. Lamoreaux, A. Lindner, and K. A. van Bibber, \arnps{65}{485}{2015}{Experimental searches for axion and axion-like particles}.


\bibitem{fed19}
M. A. Fedderke, P. W. Graham, and S. Rajendram, \prd{100}{015040}{2019}{Axion dark matter detection with CMB polarization}.
\bibitem{co19}
R. T. Co, A. Pierce, Z. Zhang, and Y. Zhao, \prd{99}{075002}{2019}{Dark photon dark matter produced by axion oscillations}.
\bibitem{sen18}
S. Sen, \prd{98}{103012}{2018}{Plasma effects on lasing of a uniform ultralight axion condensate}.
\bibitem{ros18}
J. G. Rosa and T. W. Kephart, \prl{120}{231102}{2018}{Stimulated axion decay in superradiant clouds around primordial black holes}.



\bibitem{herd14}
C. A. R. Herdeiro and E. Radu, \prl{112}{221101}{2014}{Kerr Black Holes with Scalar Hair}.

\bibitem{wal74}
R. M. Wald, \prd{10}{1680}{1974}{Black hole in a uniform magnetic field}.









\bibitem{shaikh18} 
R. Shaikh, \prd{98}{024044}{2018}{Shadows of rotating wormholes}.
\bibitem{nedkova13} 
P. G. Nedkova, V. Tinchev and S. S. Yazadjiev, \prd{88}{124019}{2013}{Shadow of a rotating traversable wormhole}.
\bibitem{abdujabbarov16} 
A. Abdujabbarov, B. Juraev, B. Ahmedov, and Z. Stuchlik, \aphss{361}{226}{2016}{Shadow of rotating wormhole in plasma environment}.
\bibitem{harko09} 
T. Harko, Z. Kovacs, and F. Lobo, \prd{79}{064001}{2009}{Thin accretion disks in stationary axisymmetric wormhole spacetimes}.
\bibitem{bambi13} 
C. Bambi, \prd{87}{084039}{2013}{Broad K$\alpha$ iron line from accretion disks around traversable wormholes}.

\bibitem{matlabnum} Lloyd N. Trefethen, \emph{Spectral methods in MATLAB}, SIAM, Philadelphia, 2000





















\end{thebibliography}
\end{document}